\documentclass{aa}  
\usepackage{graphicx}
\usepackage{physics,amsmath}
\usepackage{txfonts}
\newcommand{\q}[1]{`#1'}
\usepackage{natbib}
\bibpunct{(}{)}{;}{a}{}{,} 
\usepackage[colorlinks=true, allcolors = blue]{hyperref}

\hypersetup{
  colorlinks   = true,
  urlcolor     = blue,
  linkcolor    = blue,
  citecolor   = blue 
}
\usepackage{xcolor}

\newcommand{\oc}[1]{\textcolor{black}{#1}}

\begin{document}

   \title{2D unified atmosphere and wind simulations of O-type stars}
   \author{
          D. Debnath,
          J. O. Sundqvist,
          N. Moens,
          C. Van der Sijpt,          
          O. Verhamme,
          \and
          L. G. Poniatowski 
          }

   \institute{Instituut voor Sterrenkunde, KU Leuven, Celestijnenlaan 200D, 
              3001 Leuven, Belgium,\\
              \email{dwaipayan.debnath@kuleuven.be}}

   \date{Received 9 October 2023; Accepted 13 January 2024}

   \titlerunning{2D unified atmosphere and wind simulations of O stars}
   \authorrunning{D.Debnath et al.}

  \abstract
   {Massive and luminous O-type star (O star) atmospheres with winds have been studied primarily using one-dimensional (1D), spherically symmetric, and stationary models. However, observations and theory have suggested that O star atmospheres are highly structured, turbulent, and time-dependent. As such, when making comparisons to observations, present-day 1D modeling tools require the introduction of ad hoc quantities such as photospheric macro- and microturbulence, wind clumping, and other relevant properties.}
   {We present a series of multi-dimensional, time-dependent, radiation-hydrodynamical (RHD) simulations for O stars that encapsulate the deeper sub-surface envelope (down to $T \sim 450$ kK), as well as the supersonic line-driven wind outflow in one unified approach. Our overarching aim is to develop a framework that is free from the ad-hoc prescriptions that plague present-day 1D models. Here, we start with an analysis of a small set of such multi-dimensional simulations and then compare them to atmospheric structures predicted by their 1D counterparts.} 
   {We performed time-dependent, two-dimensional (2D) simulations of O star atmospheres with winds using a flux-limiting RHD finite volume modelling technique. Opacities are computed using a hybrid approach combining tabulated Rosseland means with calculations (based on the Sobolev approximation) of the enhanced line opacities expected for supersonic flows. The initial conditions and comparison models were derived using similar procedures as those applied in standard 1D stationary model atmosphere with wind codes.}
   {Structure starts appearing in our simulations just below the iron-opacity peak at $\sim 200 $ kK. Local pockets of gas with radiative accelerations that exceed gravity then shoot up from these deep layers into the upper atmosphere, where they interact with the line-driven wind outflow initiated around or beyond the variable photosphere. This complex interplay creates large turbulent velocities in the photospheric layers of our simulations, on the order of $\sim 30-100 \ \rm km/s$, with higher values for models with higher luminosity-to-mass ratios. This offers a generally good agreement with observations of large photospheric \q{macroturbulence} in O stars. When compared to 1D models, the average structures in the 2D simulations display less envelope expansion and no sharp density-inversions, along with density and temperature profiles that are significantly less steep around the photosphere, and a strong anti-correlation between velocity and density in the supersonic wind. Although the wind initiation region is complex and highly variable in our simulations, our average mass-loss rates agree well with stationary wind models computed by means of full co-moving frame radiative transfer solutions.}
   {The different atmospheric structures found in 2D and 1D simulations are likely to affect the spectroscopic determination of fundamental stellar and wind parameters for O stars as well as the empirical derivation of their chemical abundance patterns. To qualitatively match the different density and temperature profiles seen in our multi-dimensional and 1D models, we need to add a modest amount of convective energy transport in the deep sub-surface layers and a large turbulent pressure around the photosphere to the 1D models.}

   \keywords{ Stars: massive – Stars: atmospheres - Stars: winds, outflows - Methods: numerical - Hydrodynamics - Instabilities}

   \maketitle
   
\section{Introduction}

Due to the influence of a strong line-driven outflow \citep{cak1975}, atmospheric models of massive O-type stars (O stars) are required to treat the sub-surface layers and the overlying wind simultaneously \citep{gabler_1989}. While such \q{unified} model atmosphere tools have since reached a high level of maturity to date, they have all been computed under the assumption of a 1D, spherically symmetric, and stationary atmosphere (e.g. {\fontfamily{qcr}\selectfont FASTWIND}, \citealt{santolaya, puls_2005, puls_2020, sundqvist_2018b};
{\fontfamily{qcr}\selectfont CMFGEN}, \citealt{hillier_1998, hillier_2001}; {\fontfamily{qcr}\selectfont POWR}, \citealt{grafener_2002,hamann_2004,sander_2012}). On the other hand, observations and theory strongly suggest that the coupled envelopes, atmospheres, and winds of massive O stars are highly structured and variable, involving convective and radiative envelope and wind instabilities \citep{hearn_1972, owocki_1984, glatzel_1994, bs03, cantiello_2009, Jiang_2015} as well as various observed phenomena such as photospheric \q{macroturbulence} \citep[][]{conti_1997, howarth_1997, simon_2017} and \q{wind clumping} \citep[][]{eversberg_1998, puls_2006, oskinova_2007, sundqvist_2010, surlan_2012}.  

Recent time-dependent, radiation-hydrodynamical (RHD) simulations of the so-called \q{iron-opacity peak} region \citep[][]{iglesias_1992, slothers_1993} below the surface of massive stars have indeed displayed such structured and turbulent stellar envelopes \citep[][]{Jiang_2015,jiang_2018}. However, since these models did not treat the effects of enhanced line-opacity in a medium with significant velocity gradient (the \q{line-driving} effect, \citealt{cak1975}), they could not simulate the launch of the supersonic wind outflow from around the stellar surface. In \citet{luka_2022} we developed an opacity-formalism able to treat the sub-surface layers and the line-driven wind in a unified way and then applied this in multi-dimensional simulations of the atmospheres and outflows of classical Wolf-Rayet stars \citep{nico_2022b}. 

In this work, we present the first results from our subsequent attempts to build a unified multi-dimensional atmosphere and wind modeling framework for O stars. To do so, we started from typical 1D and stationary atmospheric and wind models, extended these downwards to well below the critical iron-opacity peak region, and then perturbed them and used them as initial conditions to solve the time-dependent RHD equations. We use the {\fontfamily{qcr}\selectfont MPI-AMRVAC} tool \citep{xia_2018, rony_2023}, which solves partial differential equations (PDE) using a finite volume solver under the parallelised {\fontfamily{qcr}\selectfont MPI} framework. We applied the RHD module of {\fontfamily{qcr}\selectfont MPI-AMRVAC} \citep[][]{nico_2022a}, and used the \q{Munich Atomic Database} \citep{pauldrach_1998, puls_2000} to calculate self-consistent line-opacities following \citet{luka_2022}. In this paper, our analysis is focused on three 2D simulations of prototypical O stars in the Galaxy, including first comparisons to the sub-surface and atmospheric structures of standard 1D models. In follow-up work, we plan to use our new framework for extension toward full 3D simulations, to make various improvements to our basic modeling tools, for calculation of synthetic spectra, to further benchmark and improve corresponding 1D tools, and to perform direct comparisons to observations.  

In Sect. \ref{methods} we go through the basic RHD equations and opacity-formalism that have been used in our modeling as well as describe the 1D initial conditions along with our prescription of perturbations. In Sect. \ref{simulation_results} we present our simulation results. We explore the formation of structures in our prototypical O star models, discuss their general atmospheric and wind properties, as well as make comparisons between the different simulations. In Sect. \ref{1d_comparison}, we then compare the average properties of our 2D simulations with comparable 1D O star models. In Sect. \ref{discussion}, we discuss our main results and further improvements to our methods. Section \ref{summary} concludes our work with a summary and a future outlook.

\section{Simulation methods}
\label{methods}

\subsection{Radiation-hydrodynamics} 

The previous studies by \citet{nico_2022a, nico_2022b} developed radiation-hydrodynamic (RHD) modules of the general hydrodynamics code {\fontfamily{qcr}\selectfont MPI-AMRVAC} \citep{xia_2018} and used these to study multi-dimensional Wolf-Rayet stellar atmospheres and winds. Following their model and study, we also solve numerically the time-dependent RHD equations on a finite volume mesh with a \q{box-in-a-star}  setup, including corrections for spherical divergence terms \citep{sundqvist_2018, nico_2022a, nico_2022b}. We provide the further below. This process involves solving the mass, momentum, and energy equations for the gas:
\begin{align}
   & \partial_t \rho + \nabla \cdot (\rho \vb*{\varv}) = 0, \label{Eq:RHD1} \\
   & \partial_t (\vb*{\varv} \rho) + \nabla \cdot (\vb*{\varv} \rho \vb*{\varv} + p_{\rm g}  \mathbb{I}) = - \vb*{f}_{\rm grav} + \vb*{f}_{\rm rad}, \label{Eq:RHD2} \\
   & \partial_t e_{\rm g} + \nabla \cdot (e_{\rm g} \vb*{\varv} + p_{\rm g} \vb*{\varv}) = - \vb*{f}_{\rm grav} \cdot \vb*{\varv} +\vb*{f}_{\rm rad} \cdot \vb*{\varv} + \Dot{q}. \label{Eq:RHD3}
\end{align}
Here, $\vb*{\varv}$ is the velocity of the gas, $\rho$ is the gas density, $p_{\rm g}$ is the gas pressure, $ \mathbb{I}$ is a unit tensor, and $e_{\rm g}$ is the total gas energy which comprises internal and kinetic components given by, 
\begin{equation}
    e_{\rm g} = \frac{p_{\rm g}}{\gamma_{\rm g} -1} + \frac{\rho \vb*{\varv}^2}{2}, 
    \label{Eq:gasenergy}
\end{equation}
where we neglect any ionisation effects in the equation of state of the gas by assuming a constant adiabatic index $\gamma_{\rm g} = 5/3$. The source terms on the right-hand side of Eq. (\ref{Eq:RHD2}), $\vb*{f}_{\rm grav}$ and $\vb*{f}_{\rm rad}$, are the gravitational and radiation force respectively. The heating and cooling of the gas due to its interaction with radiation is given by $\Dot{q}$. Since $\Dot{q}$ and $\vb*{f}_{\rm rad}$ are dependent on the radiation field, we need to treat radiation and its coupling with the gas formally. To do so, we write the energy equation for the radiation field\footnote{Here, this is the frequency-integrated $0^{\rm th}$ angular moment of the time-dependent radiative transfer equation in the comoving frame} as: 
\begin{equation}
 \partial_t E_{\rm rad} + \nabla \cdot (E_{\rm rad} \vb*{\varv} + \vb*{F}_{\rm diff}) = - \Dot{q} - \nabla \vb*{\varv} : \vb*{P_{\rm rad}}. 
 \label{Eq:Rad_energy}
\end{equation} 
Here, $E \equiv (4 \sigma/c) T_{\rm rad}^4 = a_r T_{\rm rad}^4  $  represents the frequency-integrated radiation energy density for a radiation temperature $T_{\rm rad}$. Also, $\vb*{P_{\rm rad}}$ is the frequency-integrated radiation pressure tensor and  $a_r$ is the radiation constant; $E$ is related to the mean intensity $J$ as $E = (4 \pi/c) J$ and to the radiation pressure via the Eddington tensor $\vb*{f} = \vb*{P_{\rm rad}}/E$.
Additionally, $\nabla \vb*{\varv}: \vb*{P_{\rm rad}}$ on the right-hand side of Eq. (\ref{Eq:Rad_energy}) is the dyadic product between the gradient of the velocity vector and $\vb*{P_{\rm rad}}$. 
\oc{In our formulation, the conservation equations above 
are solved on a Cartesian grid as the multigrid method implemented to solve the diffusive part of the radiation equation is not adapted for spherical meshes \citep{janis_rony_19, nico_2022b}. As such, in the present extended simulations, we need to modify certain terms to account for sphericity effects on the fluxes. Specifically, a direction (here $x$) in the Cartesian setup is taken to be the corresponding radial ($r$) direction, and terms like the divergence operator in this direction are then modified to account for sphericity by adding appropriate source terms to the conservation equations. The extent of the tangential directions (here only one) is then assumed to be small so that effects of curvature can be neglected and fluxes in the tangential directions do not need to be corrected. The method thus effectively becomes equivalent to solving the equations on a spherical mesh, but neglecting curvature effects within a small interval in the tangential direction. The exact source terms that are added for individual equations are given in Appendix A of \citet{nico_2022b}.} 

The radiative flux $\vb*{F}_{\rm diff}$ is derived from 
\begin{equation}
    \vb*{F}_{\rm diff} = - \frac{c \lambda}{\kappa \rho} \nabla E = - D \nabla E ,
    \label{Eq:fluxlimtter}
\end{equation}
where $\kappa$ is the flux-weighted opacity, $c$ the speed of light, and $\lambda$ a flux-limiter obtained from an approximation that bridges the analytic and opposite optically thick and free-streaming limits, also providing the required Eddington tensor, and the second equality introduces the radiative diffusion coefficient $D$ \citep[see Eq. (13) - (20) and discussion in][]{nico_2022a}. The radiative force density is 
\begin{equation}
    \vb*{f}_{\rm rad} = \rho \frac{\kappa \vb*{F}_{\rm diff}}{c}. 
    \label{Eq:g_rad}
\end{equation}
As in \citet{nico_2022a, nico_2022b}, throughout this paper, we assume the flux-, Planck- and energy-weighted mean opacities to be the same, \oc{given by Eq. \ref{Eq:kappa} below}. Thus 
\begin{equation}
    \Dot{q} = c \kappa \rho E - 4 \pi \kappa \rho B ,
    \label{Eq:heating}
\end{equation}
where $B \equiv \sigma T_g^4/\pi$ is the frequency-integrated Plank function for gas temperature $T_g = p_g \mu m_H/(k_b \rho)$. We assume a constant mean molecular weight $\mu = 0.61$ and our method allows for different radiation and gas temperatures. $k_b$ is the Boltzmann constant and $m_H$ is the mass of a proton. Here, the gravitational acceleration is described by: 
\begin{equation}
    \vb*{f}_{\rm grav} = \rho \frac{G M_\star}{r^2} \vb*{\hat{r}} = g_{\rm g} \vb*{\hat{r}},
    \label{Eq:grav}
\end{equation}
meaning we have neglected a small modeled envelope and wind mass in comparison to the underlying stellar mass $M_\star$ (a good assumption for the O stars considered here).

\subsection{Hybrid opacity \& finite disk correction factor}
\label{hybrid_opacity_sec}

To account properly for the \q{line-driving} \citep{cak1975} effect in the moving parts of O star atmospheres and winds, we use the \q{hybrid opacity} technique introduced by \citet{luka_2021} \oc{(see also Appendix A for connection to \citet{cak2004}'s \q{expansion opacity})}. 
\oc{This approximates} the total opacity of the medium as the sum of:
\begin{equation}
    \kappa = \kappa^{\rm Ross} + \kappa^{\rm line}.
    \label{Eq:kappa} 
\end{equation}
Here, $\kappa$ is the total opacity of the medium, $\kappa^{\rm line}$ is the opacity due to lines in a supersonic medium (line-driving effect), and $\kappa^{\rm Ross}$ are the Rosseland mean opacities 
(in this paper $\kappa^{\rm Ross} = \kappa^{\rm OPAL}$ as taken from tabulations by the \q{OPAL} project, \citealt{opal}). 

Following \cite{luka_2022}, we compute $\kappa^{\rm line}$ under the \citet{sobolev_1960} and local thermodynamical equilibrium (LTE) approximations, summing over all lines present in the \q{Munich Atomic Database} (in total $\sim 4 \times 10^6$ lines, \citealt{pauldrach_1998, pauldrach_2001}), such that 
 \begin{equation} 
	\kappa^{\rm line} = f^{*} \kappa_0 \sum_{\rm i} w_{\rm \nu,i} q_{\rm i} 
	\left( \frac{1- e^{-q_{\rm i} \tau}}{q_{\rm i} \tau} \right) =  f^{*}
         \kappa_0 M(\tau),
	\label{Eq:msum} 
\end{equation}
where $q_{\rm i}$ is the line-strength, $w_{\rm \nu,i}$ the illumination function (here as defined by Eqs. (11) and (12) in \citealt{luka_2022}) and the index $i$ runs over all lines in the database. Then we have:
\begin{equation}
    \tau = \kappa_0 c \rho {\left| \frac{dv_r}{dr} \right|}^{-1}
    \label{Eq:sobolev} 
\end{equation}
which is a characteristic Sobolev optical depth, with $\kappa_0$ as the normalisation constant. In contrast to the previous Wolf-Rayet simulations of \cite{nico_2022a} we here also introduce the so-called finite disk correction factor, $f^{*}$ (see below). 

It is important to realize that while the sum in Eq. (\ref{Eq:kappa}) typically reproduces the correct optically thick and thin limits, it is only an approximation in between. If the medium becomes very dense or the velocity gradient approaches zero or both, $\tau \, \, (\propto \rho/|d \varv_r/dr|)$ becomes very large and $\kappa^{\rm line} \rightarrow 0$, so that $\kappa \rightarrow \kappa^{\rm Ross}$. On the other hand, in the limit that all lines would be optically thin, the sum in Eq. (\ref{Eq:msum}) yields $\kappa^{\rm line} (0 < q_i \tau \ll 1) \rightarrow \sum_i w_{\nu,i} q_i \equiv \bar{Q}$, where we have introduced \citet{gayley_1995}'s $\bar{Q}$ (also see below). Following \citet{luka_2022} this becomes $\bar{Q} = \sum_i B_{\nu,i} \alpha_{l,i} / (B \rho)$, where $\alpha_l$ is the frequency-integrated line extinction coefficient. Additionally, in such diluted parts of our simulations, we typically find that the OPAL opacities approach the Thomson opacity $\kappa^{\rm Ross} \rightarrow \kappa^{\rm Th}$. That is, in this limit $\kappa$ approaches an appropriate Planck mean for an ensemble of (intrinsically non-overlapping) lines and a Thomson scattering continuum. In an \q{intermediate} region, however, lines could in principle be \q{double-counted} as they may contribute in both $\kappa^{\rm line}$ and $\kappa^{\rm Ross}$ (see also Fig. 2). This is potentially a significant weakness in our current opacity approximation, which should be further investigated in future work. On the other hand, we indeed find good agreement between average mass-loss rates of the simulations presented here and those computed for O stars by means of full frequency-dependent comoving-frame radiative transfer (in 1D and assuming a steady-state), suggesting that the issue may not be critical, at least not for the parameter range explored in this paper (see further discussion in Sect. \ref{mass_loss}). Moreover, as outlined in Appendix \ref{appendix_a} our approximation is also equivalent to the modification of Sobolev-based \q{expansion opacity} formulae suggested by \citet{cak2004} based on calculations of Fe III lines. 

\begin{figure}
    \centering
    \includegraphics[width=9cm]{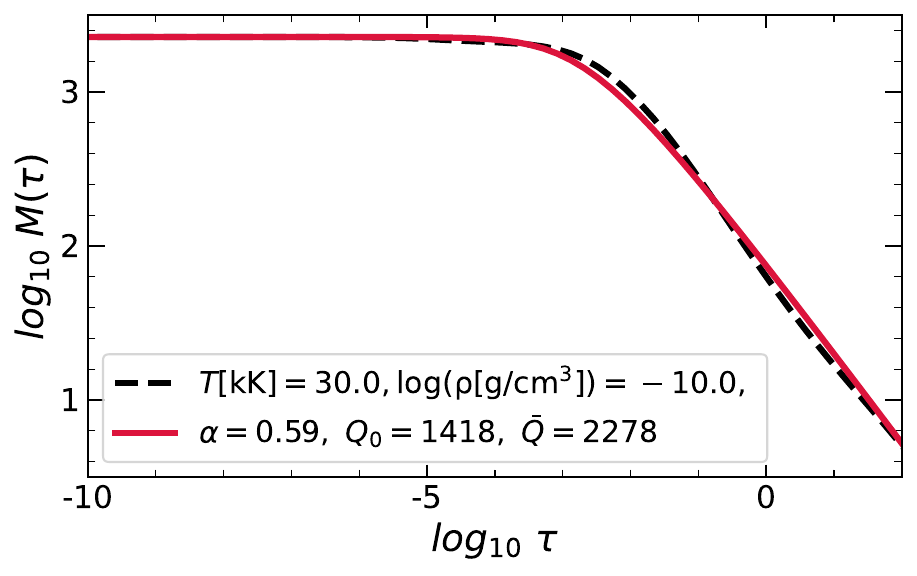}
    \caption{Figure displays the line force multiplier $M(\tau)$ as a function of the characteristic Sobolev optical depth $\tau$. The black-dashed line is plotted by calculating $M(\tau)$ at temperature $30 \rm kK$ and density $10^{-10} \ \rm g/cm^3$, whereas the red line is calculated by fitting $M(\tau)$ using $M_{\rm G} (\tau)$ in Eq. (\ref{Eq:kappaline}). The best fit for the line force parameters for this particular temperature and density are $\Bar{Q} = 2278 $, $Q_{0} = 1418 $, and $\alpha = 0.59$. }
    \label{line_force}
\end{figure}

\begin{figure}
 \centering
 \includegraphics[width=9cm]{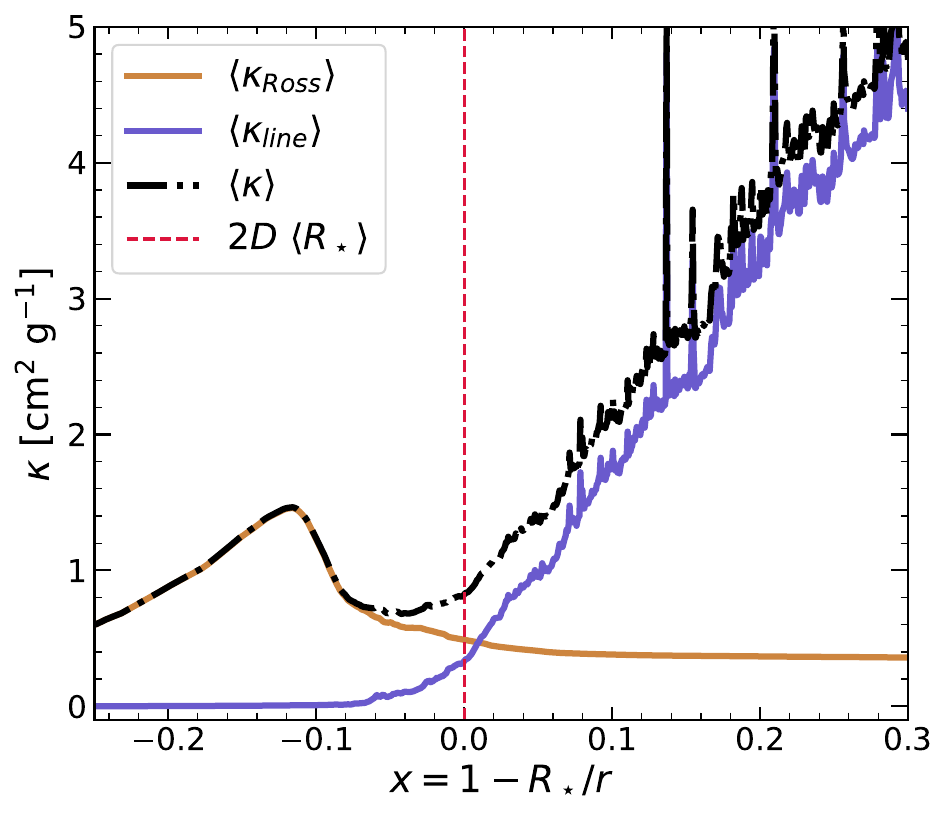}
      \caption{Figure shows averaged opacity $\langle \kappa \rangle$ as a function of scaled radius $x = 1- R_\star/r$, with $R_\star$ as the 2D averaged optical photosphere (red-dashed line). In the figure, the brown curve is the Rosseland mean opacity and the purple curve is the finite-disk corrected line-opacity. The black dash-dotted curve is the total opacity using the hybrid opacity scheme.}
      \label{hybrid_opacity}
\end{figure}

The results from the complete line-ensemble summation are fitted using \citep{gayley_1995}:  
\begin{equation}
    \kappa^{\rm line} = f^{*} \kappa_0 \frac{\bar{Q}}{1 - \alpha} \frac{\left({\left(1 - Q_0 \tau\right)}^{1 - \alpha} -1 \right)}{Q_0 \tau} = f^{*} \kappa_0 M_{\rm G} (\tau). 
    \label{Eq:kappaline} 
\end{equation}
Here, we choose the normalisation $\kappa_0 = 0.34 \ \rm cm^2 g^{-1}$, which reflects the Thomson scattering opacity in a fully ionised plasma with a chemical composition similar to that of the Sun. Also, $\bar{Q}$, $Q_0$, and $\alpha$ are the line force fit parameters; $\bar{Q}$ sets the maximum value of the line opacity in the limit where all lines are optically thin (\oc{see above}), whereas $Q_0$ and $\alpha$ give an effective maximum line strength and power-law index for a mixture of optically thick and thin lines, respectively. For this work, we have computed a large table of $\kappa^{\rm line}$ values appropriate to the conditions for the O stars under consideration. For each density and temperature pair, we use the three line force fit parameters to fit $M_G(\tau)$ to the integrated $M(\tau)$ calculated directly from our atomic database, namely, we obtain and tabulate $\alpha(\rho,T)$, $\bar{Q}(\rho,T)$, and $Q_0(\rho,T)$; an example is given in Fig. \ref{line_force}. 
These fitted and tabulated parameters are then used within the time-dependent simulations, providing varying and locally consistent values for $\kappa^{\rm line}$. We note that since the line force parameters thus vary in space and time depending on the ionisation state of the model, there is no need to introduce further fit-parameters (such as Abbott's \q{$\delta$}, see discussion in \citealt{luka_2022}). 
The technique is similar to how OPAL opacity tables are constructed from pre-calculations of the Rosseland mean opacity. Similarly to the OPAL tables, $\kappa^{\rm line}$ can be calculated for various chemical compositions. To obtain relative abundances we here use the default scale \citep{grevesse_1993} from the OPAL tables, thereby, hydrogen mass-fraction $X = 0.70$, helium mass-fraction  $Y= 1 - Z_\odot = 0.28$ for metallicity $Z_\odot = 0.02$. We note that this differs slightly from the often-used chemical abundance scale by \citet{asplund_2009} (with $Z_\odot = 0.013$). Fig. \ref{hybrid_opacity} displays \q{average} opacity curves for one of the 2D O star models presented in the next section. The figure illustrates a characteristic opacity behaviour that is dominated by the Rosseland mean in deep optically thick layers, and by the line-opacity component in the wind outflow above the stellar surface. To a large extent, the diminishing contribution of line-driving to the total opacity in deep atmospheric layers is simply due to the inverse dependence with density seen in Eq. (\ref{Eq:kappaline}).     

As shown above, $f^{*}$ is the standard finite-disk correction factor to the line force \citep{pauldrach_1986, friend_1986}, as given by, for instance, Eq. (29) in \citet{luka_2022}, \oc{which corrects the radial streaming expression by taking into account non-radial light rays and velocity gradients. It is here} numerically implemented as in \citet{puls_owocki_1996}. Above the stellar surface, this produces a varying $f^{*}$ depending on the local wind conditions. 
In the deeper regions of the atmosphere, we set a constant $f^{*} = 1 /(1 + \alpha_{\star})$ (the limiting value at the stellar surface), where $\alpha_{\star}$ is the line force parameter $\alpha$ at the photosphere $R_\star$ (see definition below). Since $R_\star$ is not constant in our models (instead it varies with time depending on the local conditions in the atmosphere), in practice, we let our models relax until the overall variation of the stellar radius is small and then we use a constant $R_\star$ in the evaluation of $f^{*}$ for the remainder of the simulation.
As can also be seen in Fig. \ref{hybrid_opacity}, the values of $\kappa^{\rm line}$ are mostly small as compared to $\kappa^{\rm Ross}$ in the layers below $R_\star$. A fixed $R_\star$ for the computation of $f^{*}$ is nevertheless used here in order to avoid additional initiation of structures due to small variations of $f^{*}$ at the stellar surface, as our implementation is approximate anyway and used here primarily to avoid issues related to \q{over-loaded} line-driven wind solutions in the radial streaming case. Indeed, strictly speaking, $f^{*}$ as implemented here is valid only for spherically symmetric systems; a full multi-dimensional evaluation would require costly numerical integration across the stellar disk (see, e.g. \citealt{petrenz_puls_2000, kee_2016}). Notwithstanding this caveat, this 1D finite-disk correction factor has been quite successfully applied in a range of other non-spherical line-driven systems like magnetically channeled O star winds \citep{ud_doula_2002}. Nevertheless, this approximate 1D treatment should be improved upon in future work.

\subsection{Initial conditions}
\label{Initial_conditions}

\begin{figure}
 \centering
 \includegraphics[width=9cm]{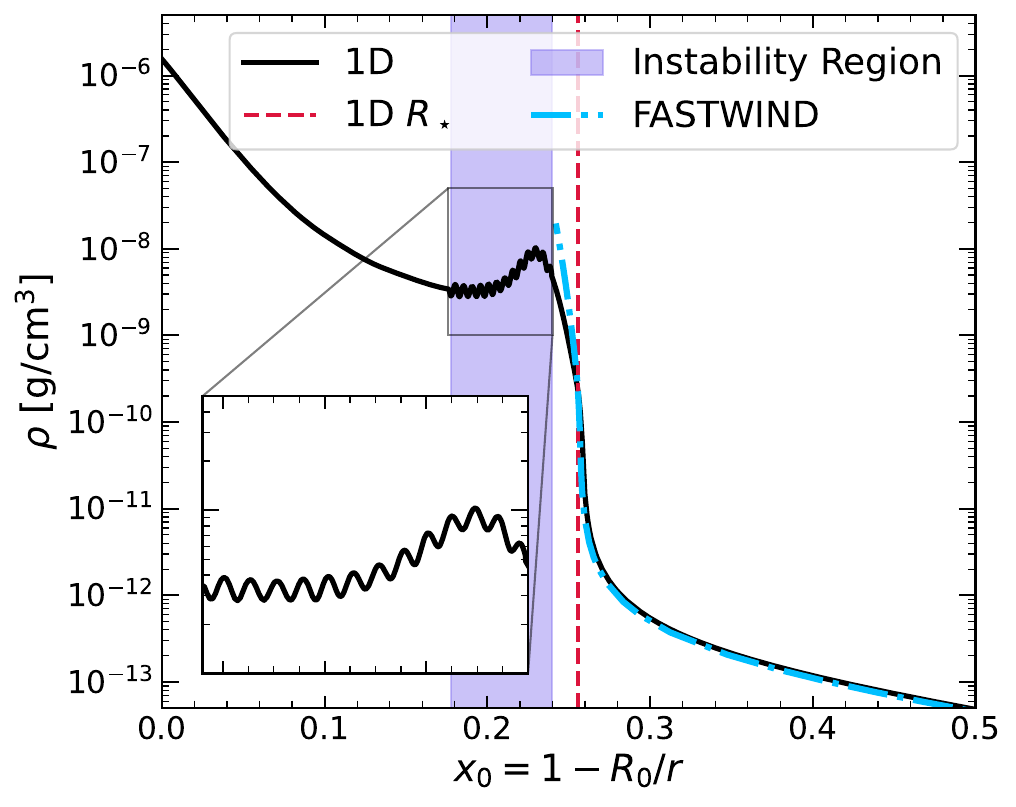}
      \caption{Initial 1D  gas density [in $g/cm^3$] structure as a function of modified radius $x_0 = 1- R_0/r$, with $R_0$ as the lower boundary radius. The cyan curve is corresponding {\fontfamily{qcr}\selectfont FASTWIND} density structure as described in Sect. \ref{Initial_conditions}. The red-dashed vertical line is the optical photosphere $R_\star$. The shaded-purple region is the instability region where we provide the initial perturbations (see zoomed-in part).}
     \label{perturbed_density}
\end{figure}

To compute the initial conditions for our time-dependent simulations, we used a procedure similar to that outlined in Sect. 2 of \citet{santolaya} ($1^{st}$  version of the {\fontfamily{qcr}\selectfont FASTWIND} unified model atmosphere code). This method assumes a spherically symmetric, stationary, and analytic wind structure described by a radial \q{$\beta$} velocity law $\varv_r(r) = \varv_\infty (1-b r_{\rm T}/r)^\beta$ and mass density $\rho = \dot{M}/(4 \pi r^2 \varv_r)$ on top of an atmosphere in (quasi-)hydrostatic and (here) local thermal equilibrium. The two components of the atmosphere are smoothly connected at a transition velocity $\varv_{\rm T} = f c_{\rm iso, gas} (T_{\rm eff})$, where $c_{\rm iso,gas} = \sqrt{\frac{p_{\rm g}}{\rho}}$ is the isothermal gas sound speed, the parameter $f \approx 0.1-1$, and $b$ is derived from $\varv_{\rm T}$; this analytic wind structure provides an intermediate boundary condition for inward integration of the static gas and radiative momentum equations on a suitable grid of column mass $m$. For a given set of input parameters $L_\star$, $M_\star$, $R_\star$, elemental abundances, $\dot{M}$, $\varv_\infty$, $\beta$ (=1 throughout this work) then, this procedure is iterated until a unified structure that fulfills 
\begin{equation}
R_\star \equiv r\,(\tau_R=2/3) 
 \label{Eq:Rstar}  
\end{equation}
is found, where $\tau_R$ is the spherically modified Rosseland optical depth (see below). In practice, for opacity, $\kappa$, and gas pressure, $p_{\rm g}$, we solve numerically the following coupled differential equations in the layers below the intermediate boundary given by $r_T$:  
\begin{align}
    & \frac{dp_{\rm g}}{dm} = 
    g_\star \left(1- \Gamma \right) \left( \frac{R_\star}{r} \right)^2, 
    \label{Eq:HE} \\
    & \frac{dT}{dm} = \frac{3 \kappa T_{\rm eff}^4}{16 T^3} 
    \left( \frac{R_\star}{r} \right)^2, \\
    & \frac{d \tau_R}{d m} = \kappa  \left( \frac{R_\star}{r} \right)^2, \\
    &    \frac{dr}{dm} = -\frac{1}{\rho},
\end{align}
along with the definitions and relations 
\begin{align}
    & \sigma T_{\rm eff}^4 \equiv F_\star \equiv \frac{L_\star} {4 \pi R_\star^2}, \label{teff}  \\
    & g_\star \equiv \frac{ G M_\star}{R_\star^2}, \label{gstar} \\
    & \Gamma \equiv \frac{g_r}{g_g} = \frac{L_\star \kappa}{4 \pi G M_\star c}, \\
    & p_{\rm g} = \frac{\rho k_b T}{\mu m_H},
\end{align}
where $g_r$ is the radial component of the radiation force as seen in Eq. (\ref{Eq:g_rad}). Also, for simplicity, we assume a constant $\mu=0.61$ (set by assuming fully ionised hydrogen and helium). Above, 
it is further implicitly assumed that the gas and radiation temperatures are equal at $T$. We also note that since here $dP_{\rm rad}/dm = g_{\rm r}/\rho = \kappa F/c = \kappa L_\star/(4 \pi r^2 c)$, Eq. (\ref{Eq:HE}) above may equivalently be formulated as a hydrostatic equation for the total pressure $P_{\rm tot} = P_{\rm rad} + p_{\rm g}$ of the radiating gas. To classify models, it is further useful to introduce the classical Eddington parameter, $\Gamma_e$, and luminosity, $L_{\rm edd}$: 
\begin{equation} 
    \Gamma_e \equiv \frac{L_\star \kappa_e}{4 \pi G M_\star c} \equiv \frac{L_\star}{L_{\rm edd}},  
\end{equation}
where $\kappa_e$ is the electron scattering opacity for a fully ionised medium of a certain chemical composition.

Two differences between our procedure here to compute initial conditions and the method outlined in \citet{santolaya} are that we (i) do not account for any variations in \q{Hopf-parameters}, but instead assume the Eddington approximation $3 P_{\rm rad} = E_{\rm rad} = a_r T^4 $ is valid throughout the atmosphere and (ii) use opacities $\kappa = \kappa^{\rm Ross}$ from the OPAL database (instead of fitting a Kramer's like law) for a given set of input chemical abundances. The latter point allows us to probe much deeper atmospheric layers than {\fontfamily{qcr}\selectfont FASTWIND} so that we are able to consider the critically important regions around the so-called \q{iron-opacity peak} at $T \sim 2 \times 10^5$ K; specifically, while a typical {\fontfamily{qcr}\selectfont FASTWIND} model only extends down to column masses $m \sim 100$ we here terminate the downward integration only when a high temperature $T \sim 4-5 \times 10^5$ K has been reached (which in the O star regime occurs at $m \gg 100$). 

The black curve in Fig. \ref{perturbed_density} indicates the density structure for a template \q{early} O star model (with small initial perturbations around the convective instability region, see Sect. \ref{perturbations}).
Additionally, the cyan curve corresponds to the {\fontfamily{qcr}\selectfont FASTWIND} model computed for the same set of input parameters. This model indeed agrees very well with our simplified 1D prescription, however the {\fontfamily{qcr}\selectfont FASTWIND} calculation only extends down to layers with $\sim 100$ kK, as explained above.\footnote{To make suitable comparisons, we have here adjusted the standard input specifications of more recent versions of {\fontfamily{qcr}\selectfont FASTWIND} \citep[][]{puls_2005, sundqvist_2018b} by opting for the \q{Lucy temperature structure} \citep{lucy_1971} without further corrections, neglecting any wind clumping effects, and choosing a solar abundance scale reflecting the calculations of this paper.} The input parameters $T_{\text{eff}}$, $g_\star$ and $\dot{M}$ for the {\fontfamily{qcr}\selectfont FASTWIND} model are the same as those listed in 
Table \ref{table:Models} (O4 model). We note that the table excludes the terminal wind speed $\varv_{\infty}$, which for the 1D model is set to $2100$ km/s, made to correspond to an extrapolation of the results of the 2D \q{average} structure, and which is also typical for observed stars at this temperature range \citep{hawcroft_2023}. 

From Fig. \ref{perturbed_density} (and later in Fig. \ref{init_density_temp}), we can further see that in the region around the above-mentioned iron-opacity peak, there is a clear density-inversion. This occurs because in this region of a high Rosseland mean opacity, $\Gamma > 1$, such that the gradient in the hydrostatic Eq. (\ref{Eq:HE}) changes sign \citep[see also e.g.][]{grafener_2012, Jiang_2015, kohler_2015, sanyal_2015}. To overcome this region of enhanced opacity the radiative stellar envelope reacts by expanding significantly, similar to what has been observed in envelope-models of more evolved massive stars \citep{petrovic_2006, grafener_2012, grassitelli_2018, luka_2021}. Moreover, the transition region to the analytic wind outflow is also clearly visible in the figure, in particular through the characteristic change from an essentially exponential atmosphere to a more modest $\sim r^{-2}$ drop-off in density.

\subsection{Numerical grid}
\label{numerical_grid}

\begin{figure}
 \centering
 \includegraphics[width=9cm]{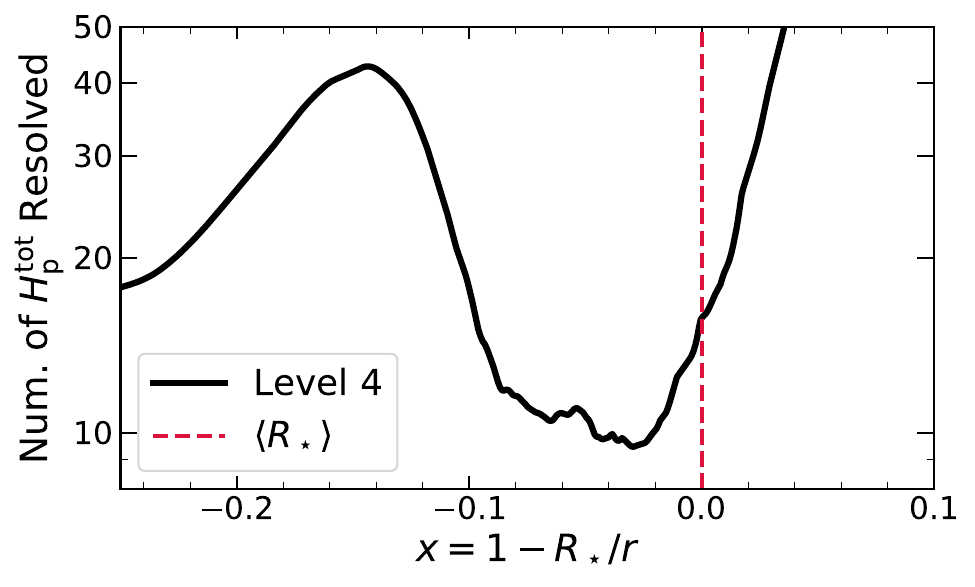}
      \caption{Number of total pressure height scale $H_{\rm p}^{\rm tot}$ resolved per cell in our simulation. Level 4 corresponds to the highest level of resolution in our models. Levels 3-1 are not shown since it is zoomed in to indicate the deep layers and the photosphere. See text for details.}
     \label{resolution}
\end{figure}
The lower boundary of our simulation is situated inside the stellar envelope, at a fixed $R_0$ given by the initial condition calculation. The outer boundary extends well into the supersonic regions of the outflow, we set it here to $r= 3 R_0$. This allows us to take a simultaneous and \q{unified} approach to study the deep and optically thick layers of the atmosphere, the wind launching region, and the supersonic outflow up until a point where a majority of the gas particles have reached escape velocities. 

The simulations that we present in this paper have been run on a Cartesian mesh, with four levels of refinement. Although {\fontfamily{qcr}\selectfont MPI-AMRVAC} is capable of adaptive mesh refinement (AMR), in our simulations, we use a fixed-radius refinement since otherwise, we run into methodological problems with small-scale \q{clumped} regions in particularly the wind outflow. At our lowest resolution (Level 1), the numerical domain is 32 cells in the x-direction and 256 cells in the y-direction covering $0.2 R_0$ and $3 R_0$, respectively. The y-direction is here identified with the radial coordinate $r$, applying the spherical correction-terms discussed in \citet{nico_2022a, nico_2022b}, and the x-direction is thus the lateral coordinate. Three times doubled to the highest resolution means we have 256 and 2048 cells in the lateral and radial directions, respectively. The highest level resolution (Level 4) cell covers $9.76 \times 10^{-4} \ R_0$ in length. This highest resolution is set from the lower boundary $R_0$ of the simulation up to the stellar surface $R_\star$ of the 1D model used as an initial condition. 

Such a refinement is very useful for achieving proper resolution in the deeper optically thick atmospheric regions without net radial outflow while still keeping the total number of cells in the outer supersonic outflow regions computationally viable. To this end, we ensured that we resolve a few pressure scale heights in our simulations, especially in the deeper regions up to $\sim R_\star$. For a radiating gas in hydrostatic equilibrium, this pressure scale height is: 
\begin{equation}
    H_{\rm p}^{\rm tot} = \frac{c_{\rm iso,gas}^2}{{g \beta_p}}, 
    \label{Eq:Htot}
\end{equation}
where $\beta_p =  p_{\rm g}/P_{\rm tot} = p_{\rm g}/\left(p_{\rm g} + P_{\rm rad}\right)$.
Fig. (\ref{resolution}) confirms a posteriori that we do indeed resolve several total pressure scale heights in layers deeper than and around $R_\star$.

\subsection{Perturbations}
\label{perturbations}

To initiate structure formation in our simulations, the initial conditions are perturbed in the predicted sub-surface convective region. This region may be identified following the linear analysis in \citet{bs03}, where instabilities in an optically thick radiation-pressure dominated gas were studied. The instability criterion for convection then becomes:
\begin{equation}
    \frac{\gamma_{\rm g} p_{\rm g}}{4(\gamma_{\rm g} - 1)E_{\rm rad}}N_g^2 + \frac{1}{3} N_r^2 < 0,
\end{equation}
where 
\begin{align*}
    N_g^2 &\coloneqq - \vb*{f}_{\rm grav} \cdot \left(\frac{1}{\rho c_{s, gas}^2} \nabla p_{\rm g} - \nabla \ln \rho \right), \\ 
    N_r^2 &\coloneqq - \vb*{f}_{\rm grav} \cdot \left(\frac{1}{3\rho c_{s, rad}^2} \nabla E_{\rm rad} - \nabla \ln \rho \right).
\end{align*}
These are the Brunt-Vaisala frequencies for the gas and radiation parts, respectively. This instability criterion may be interpreted as a modified \citet{schwarzschild_1958} criterion for media with significant radiation pressure.  $c_{s, rad}$ and $c_{s, gas}$ are the adiabatic radiation and gas sound speeds defined below in Sect. \ref{timescales}. The predicted convective instability region for the 1D initial condition model of a prototypical O star is shown in Fig. \ref{perturbed_density} as purple-shaded parts. 

In this convective region, we add small perturbations (max 20\%) to the initial density profile (see the zoomed-in part in Fig. \ref{perturbed_density}). Such perturbations have a plane wave dependence on $\mathbf{r}$, namely,
\begin{equation}
    \delta \rho \sim A \exp{i \mathbf{k} \cdot \mathbf{r}},
\end{equation}
with $A$ a constant set here to $0.2 \rho$ and $\mathbf{k}$ the wavevector of the perturbation.
The final perturbations are a superposition of several wave modes with different wave numbers $k$, and we assume the perturbations have a Gaussian distribution in Fourier space. We consider a range of wavenumbers $k$ limited by the total pressure scale height and the cell size of our simulations, namely,
\begin{equation}
    k_x \in \left[\frac{2 \pi}{H_{\rm p}^\text{tot}}, \frac{2 \pi}{L_{\text{cell},x}}\right], k_y \in \left[\frac{2 \pi}{H_{\rm p}^\text{tot}}, \frac{2 \pi}{L_{\text{cell},y}}\right].
\end{equation}
Since we follow \citet{bs03}, where results are limited to the short wavelength regime, we set the scale height as a lower limit for the wavenumbers (and therefore also the wavelength). The number of wave modes $N(k)$ with a certain wavenumber $k$ is then a linear combination of Gaussians centered around $(k_{x,c}, k_{y,c})$, $(k_{x,c}, -k_{y,c})$, $(-k_{x,c}, k_{y,c})$ and $(k_{x,c}, k_{y,c})$, respectively, with
\begin{equation}
    k_{x,c} = (k_{x,min} + k_{x,max})/2, k_{y,c} = (k_{y,min} + k_{y,max})/2.
\end{equation}
Finally, perturbations are calculated by Fourier transforming $N(k)$, so that 
\begin{equation}
    \delta \rho = 0.2 \rho \int \exp{i \mathbf{k} \cdot \mathbf{r}} N(k) d\mathbf{k}.
\end{equation}
Additionally, in this region, we add perturbations to the lateral velocity which are sinusoidal in $r$ and $y$. 

The analysis by \citet{bs03} further shows that acoustic waves can potentially be unstable around opacity bumps in a massive stellar envelope  (see also \citealt{owocki_2015}). As in \citet{Jiang_2015}, we have focused here on the convective instability, however, we note that it is possible that such acoustic waves, which essentially are local versions of \q{strange mode} instabilities \citep{glatzel_1994}, could also be driven unstable in the present O star simulations.  

\subsection{Boundary conditions}
\label{boundary_conditions}

In the lateral direction, the boundaries are periodic for all conserved quantities solved in Eq. (\ref{Eq:RHD1} - \ref{Eq:RHD3}, \ref{Eq:Rad_energy}). The mass density at the first ghost cell at the lower boundary is fixed at the value given by the initial conditions and is then extrapolated to the remaining ghost cells assuming hydrostatic equilibrium. For the gas momentum, we let it vary, extrapolating from the first active cell to the ghost cells assuming mass conservation. We set the radial component of the radiation energy $E_{\rm rad}$ by a fixed input radiative luminosity at the bottom boundary, using a finite difference formulation of Eq. (\ref{Eq:fluxlimtter}) to extract $E_{\rm rad}$ from its gradient. The gas energy $e_{\rm g}$ is then set by assuming $T_{\rm rad} = T_g$ at the lower boundary, using the ideal gas law and Eq. (\ref{Eq:gasenergy}). 

In the case of the outer boundary, the outflow is highly supersonic, and density, momentum, and gas energy are linearly extrapolated. For the radiation energy $E_{\rm rad}$, we analytically solve Eq. (\ref{Eq:fluxlimtter}) from the outer boundary radius $r_{o}$ to $r \rightarrow \infty$ by assuming constant opacity and flux-limiter, and a diffusive flux and mass density that varies according to $r^{-2}$ outside $r_o$, yielding $E_{\rm rad,o} = \frac{F_{\rm diff,o} r_{o}}{3D_{o}}$ where subscripts $o$ in the parameters signify that values are taken at the radial outer boundary. 

\begin{figure*}
 \centering
 \includegraphics[width=18cm]{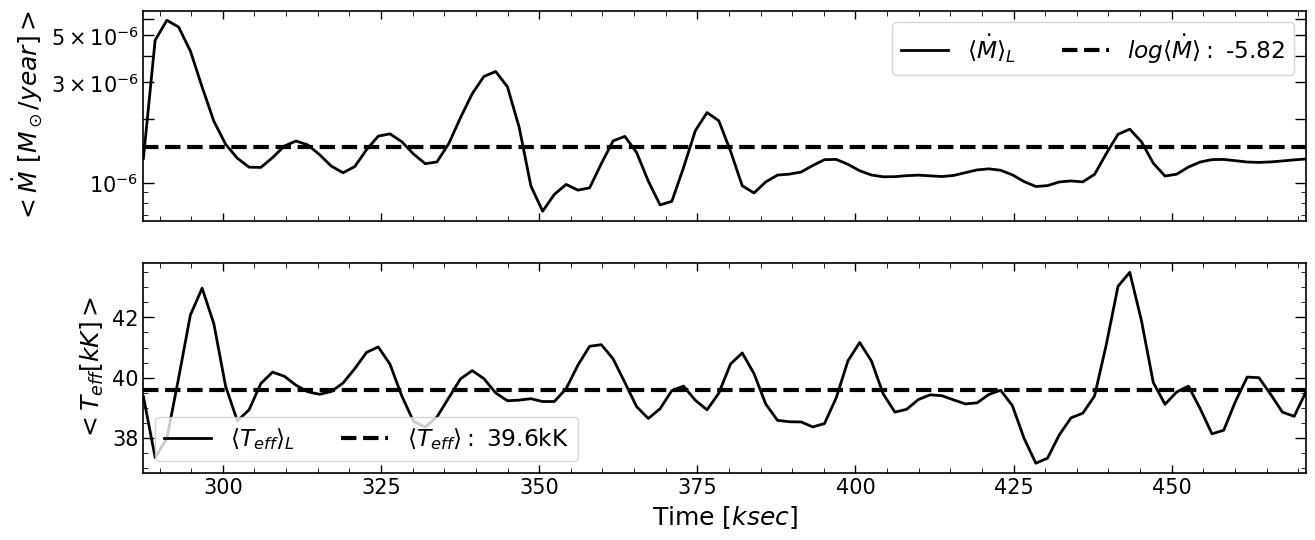}
      \caption{Mass loss rates (top) and effective temperature (bottom) for the $\rm{O} 4$ model as varying with time (shown here in $10^3$ seconds). The black curve displays the laterally averaged quantity varying across time written as $\langle 2D \rangle_L$, and the black-dashed line is the lateral-temporal averaged quantity, here $\langle 2D \rangle$.}
    \label{massloss}
\end{figure*}

\subsection{Timescales}
\label{timescales}

The dynamics of the deep, optically thick atmosphere and the outflowing wind may in general be governed by different timescales. For the deep atmosphere without a net radial outflow, we may choose a characteristic dynamical timescale, $t_d$, as: 
\begin{equation}
    t_{\rm d,a} = \frac{H_{\rm p}^{\rm tot}}{c_{\rm s}},
\end{equation} 
where $c_{\rm s}$ is the total adiabatic sound speed in the radiating gas as such, 
\begin{equation} 
    c_{\rm s}^2 = c_{\rm s,rad}^2 + c_{\rm s,gas}^2 , 
    \label{Eq:totalsound}
\end{equation} 
for adiabatic radiation and gas sound speeds of
\begin{align}
    & c_{\rm s,gas}^2 = \frac{5 p_{\rm g}}{3 \rho}, \\
    & c_{s,rad}^2 = \frac{4 P_{rad}}{3 \rho}.
    \label{Eq:radsound}
\end{align}
On the other hand, the dynamics of the outflowing wind, starting around an \q{average} sonic point $R_{\rm sonic}\simeq r \ [\varv_r > c_{\rm iso,gas}]$, which in our simulations vary in space and time depending on the local conditions, is typically better characterised by,
\begin{equation}
    t_{\rm d,w} = \frac{R_\star}{\varv} ,
\end{equation} 
with $\varv$ a characteristic wind speed; here such a typical wind speed simply has been chosen from inferences of O star winds and some visual inspection of our simulations. For the O star regime considered in this paper, typical numbers are $\varv \sim 1000$ km/s, $R_\star \sim 17 R_\odot$, $H_{\rm p}^{\rm tot} \sim 0.03 R_\star$, and $c_{\rm s} \sim 150$ km/s, yielding $t_{\rm d,a} \sim 1000$ sec and $t_{\rm d,w} \sim 10000$ sec. 

Finally, thermal relaxation of the deeper atmospheric layers by radiative diffusion proceeds on a timescale related to the total thermal energy content per unit area divided by the total energy flux. 
Following \citet{freytag_2012} this may be approximated for the case here of radiation-dominated O star atmospheres by $t_{\rm th} \sim E_{\rm rad} H_{\rm p}^{\rm tot} / F_{\rm diff} \sim 10^{3 \dots 4}$ sec. Alternatively, a thermal timescale may be estimated for the entire atmosphere following \citet[][]{grassitelli_2016, nico_2022a}, $t_{\rm th} = GM_\star\,M_{\rm atm}/\left(R_\star L_\star\right)$, 
where $M_{\rm atm}$ represents the mass of the stellar atmosphere. Integrating over the average radial density profile of our simulations from the iron-bump to the outer boundary, this gives typical numbers $\approx 40000 \ {\rm sec} \approx$ 4 $t_{\rm d,w}$. This estimate yields higher characteristic numbers than $t_{\rm d,w}$, and in practice, we thus also examine whether our simulations have reached an approximate energetic steady-state by inspecting the temporal behaviour in total luminosity, see Sect. \ref{structure_formation}.

\section{Simulation results} 
\label{simulation_results}

\subsection{Classification of models}
\label{classification}

\begin{table*}
\caption{Fundamental parameters for the $\langle 2D \rangle$ O star models studied in this paper. From left to right, the columns display the model name, effective temperature, stellar mass, radius, luminosity, Eddington ratio, surface gravity, and mass loss rates. Angle brackets denote averaged quantities as explained in the text.}       
\label{table:Models}      
\centering          
\begin{tabular}{c | c c c c c c c }  
\hline\hline       
Model & $\left<T_{\rm eff} [kK]\right>$ & $M_\star/M_\odot$ & $ \langle R_\star \rangle/R_\odot$ & $log_{10} \left(\left<L_\star\right>/L_\odot\right)$ & $\left<L_\star\right>/L_{\rm edd}$& $ log_{10} \left<g_\star\right>$ &  $log_{10} \left<\Dot{M}\right>  \ [M_\odot/yr] $ \\ 
\hline                    
   $\rm{O}8$ & 33.3 & 26.9 & 12.26 & 5.23 & 0.16 & 3.69 & -6.86 \\
   $\rm{O}4$ & 39.6  & 58.3 & 16.98 & 5.78  & 0.27 & 3.74  & -5.84\\
   $\rm{O}2$ & 43.8 & 58.3 & 15.99 & 5.93 & 0.38 & 3.79 & -5.56 \\
\hline                  
\end{tabular}
\end{table*}

\begin{figure*}
 \centering
 \includegraphics[width=18cm]{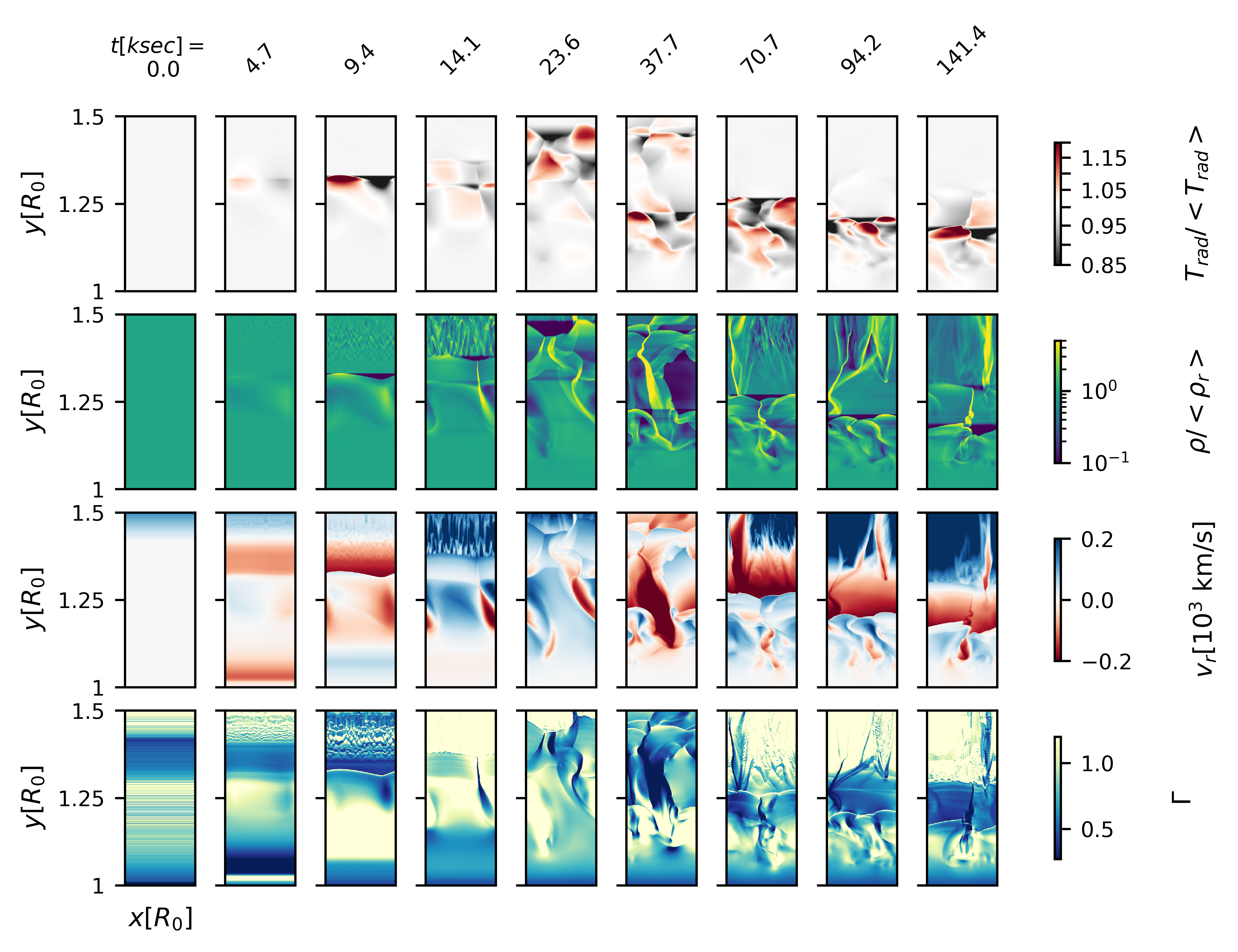}
      \caption{ Colour maps of relative radiation temperature (top panel), relative density (second panel), and the radial velocity (third panel) and $\Gamma$ (bottom panel), for several snapshots at different times (shown on top in $10^3$ seconds) at the beginning of our simulation for the O$4$ model. The figures are zoomed in to show the inner regions of our simulations. In the lateral direction, the figure extends to $0.2 \ R_0$ and in the radial direction to $1.5 \ R_0$, with $R_0$ as the lower boundary radius of our simulation.}
    \label{time_series_initial}
\end{figure*}

\begin{figure*}
 \centering
 \includegraphics[width=18cm]{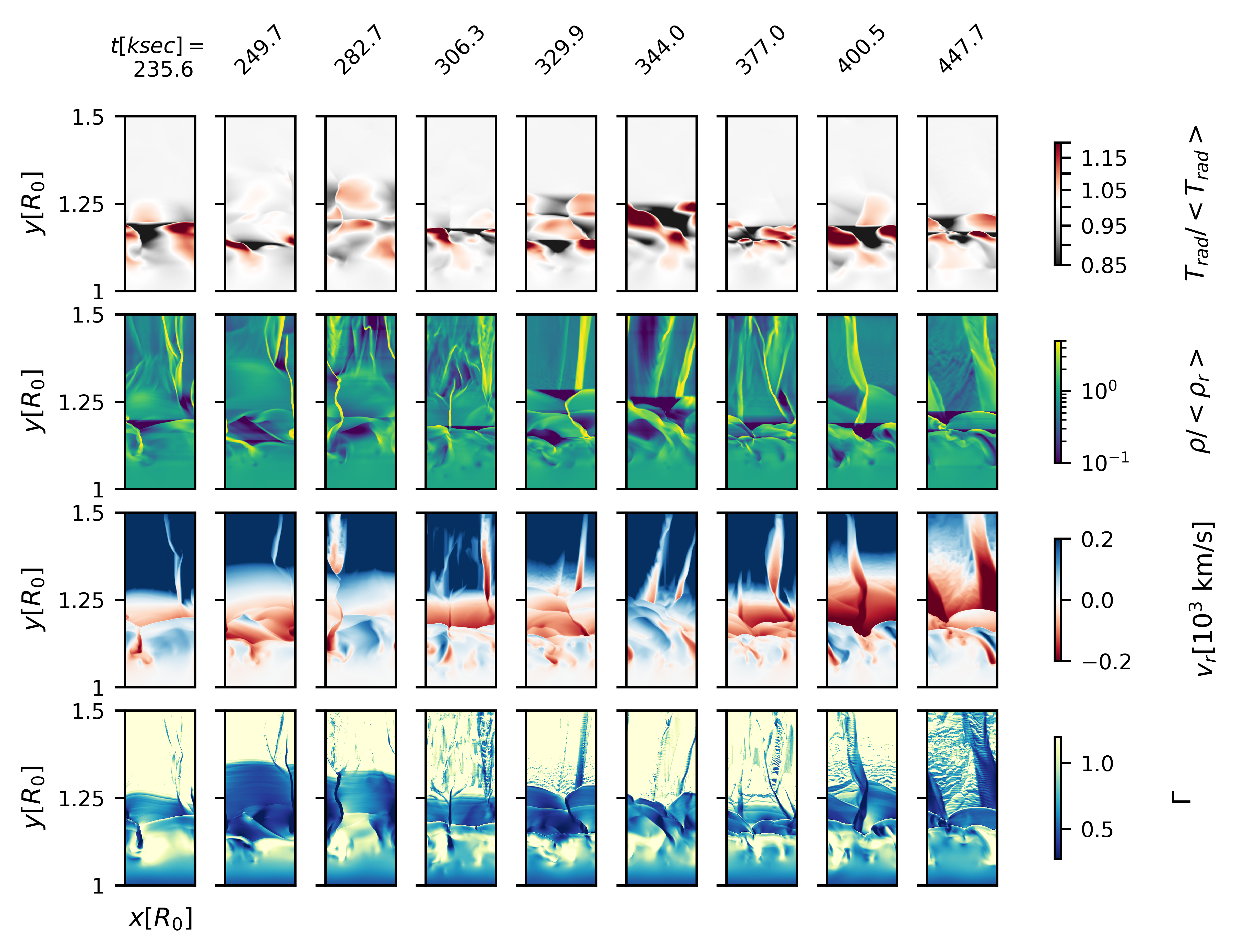}
      \caption{Similar to Fig. \ref{time_series_initial}, but now at times after the models have dynamically relaxed from their initial conditions. The black-dashed line is the averaged 2D optical photosphere $
      \langle R_\star \rangle$.}
    \label{time_series_relax}
\end{figure*}

\begin{figure}
 \centering
 \includegraphics[width=9cm]{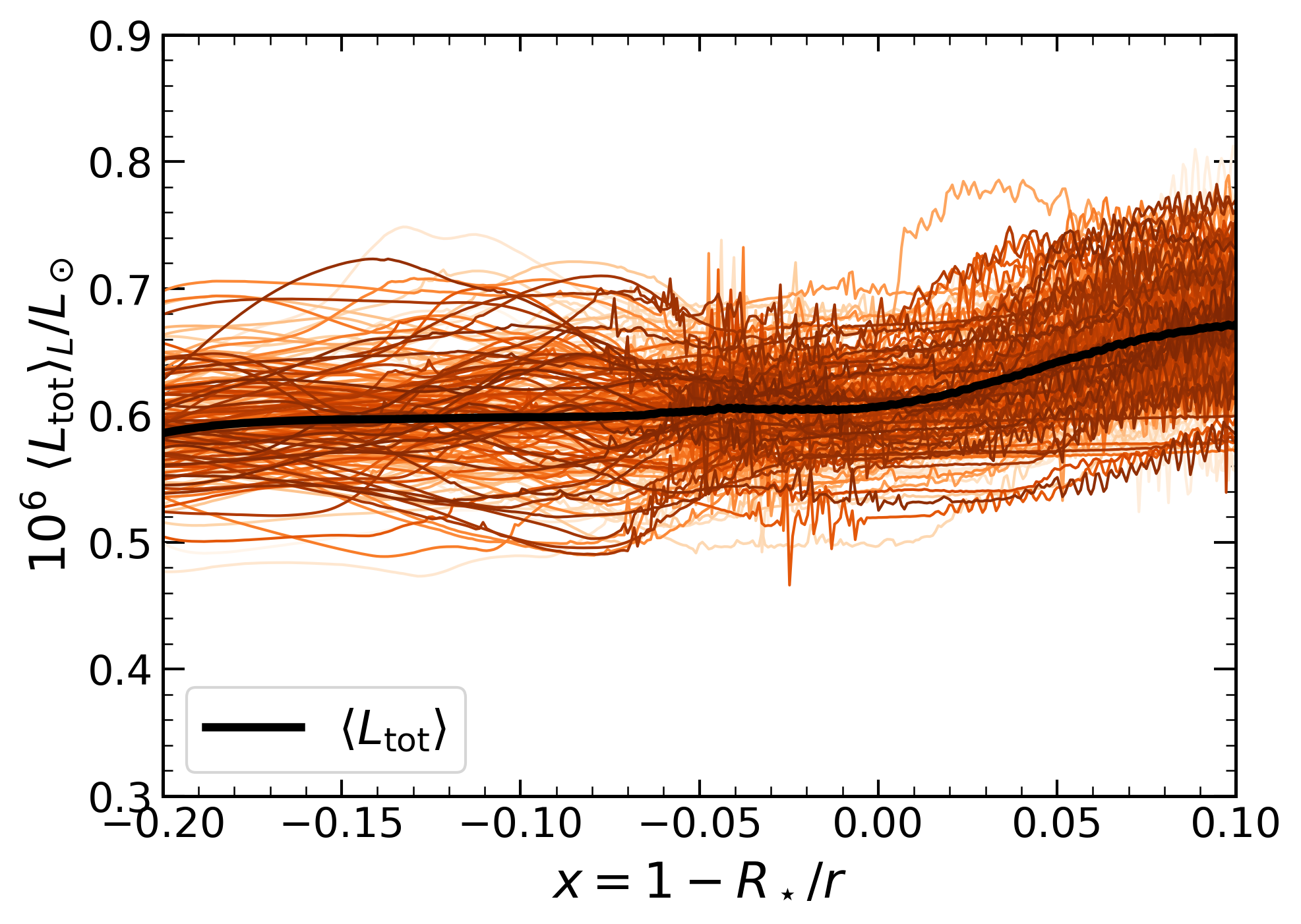}
      \caption{\oc{This figure shows running averaged total luminosity $\langle L_{tot} \rangle$ as a function of scaled radius $x = 1- R_\star/r$. The thin lines represent lateral averages taken over a few time snapshots (with lighter colours representing earlier snapshots), whereas the black line represents a long time-average of the total luminosity.}}
      \label{flux_variation}
\end{figure}
Generally, O stars are classified according to their spectral type, which is related to the effective temperature, $T_{\rm eff}$, as defined at the photospheric radius, $R_\star$ (see Eqs.  (\ref{Eq:Rstar}) and (\ref{teff})). In our 2D simulations, however, it is the mass, radius, and radiative luminosity at the lower boundary that are specified as input parameters (see Sect. \ref{boundary_conditions}). As such, the photospheric radii and effective temperatures of the simulations are emergent properties that typically vary both in time and space, and in general, they do not have the same values as the corresponding ones of the 1D initial conditions (see detailed discussion in the next section). As such, we calculate lateral and temporal averages of fundamental stellar parameters and classify our models according to them; that is, hereafter, $\langle X \rangle$ for quantity $X$ means a lateral and temporal average taken at some radial cell. In the case of $\langle L_\star \rangle$ and $\langle T_{\rm eff} \rangle$ the radial cell corresponds to $R_\star \equiv \langle r(\tau_{\rm Ross} =2/3) \rangle$, whereas for $\langle \Dot{M} \rangle$ we take an average of a few radial cells close to the outer boundary of our simulation (where gas particles have become unbound). To ensure that transients stemming from the initial conditions do not influence our results we make sure that such averages are taken several dynamical time steps after our models have dynamically relaxed. Additionally, we also make sure that we take averages only after turning off a varying $R_\star$ when computing the finite disk correction factor (see previous section).

That our models are dynamically well relaxed is also evident from Fig. \ref{massloss}, showing that for advanced simulation-times time-fluctuations in $\langle \Dot{M} \rangle$ and $\langle T_{\rm eff} \rangle$ are rather stochastic around the mean value. In the context of this Fig. \ref{massloss}, it is important to recall that we are only modeling $0.2 R_0$ of the star in one of the two lateral directions, namely, only a fraction of the overall stellar and wind volume. As such, the time-variance displayed in this figure will notably not be representative of the predicted time-variability for observations of these global stellar and wind parameters. 
Specifically, $\left<T_{\rm eff}\right>$ is obtained by computing the diffusive radiative flux at the average photosphere $\left<R_\star\right>$. This allows us to organise the first O star simulations presented here according to their $\left<T_{\rm eff}\right>$, and thereby to also name them according to their approximate \q{spectral type}. Mainly, this represents a convenient organisation of our models and, thus, it does not necessarily reflect the \q{true} spectral type the modeled stars would be classified as according to their spectral appearance. Table \ref{table:Models} provides averaged fundamental stellar parameters of our models, along with their averaged emergent mass loss rates.

\subsection{Structure formation, relaxation, and initial conditions}
\label{structure_formation}

In Figs. \ref{time_series_initial} and \ref{time_series_relax}, we show snapshots from the temporal evolution of our prototypical 
\q{O$4$} simulation during the initial and dynamically relaxed phase. As can be seen in Fig. \ref{time_series_initial}, structure starts developing within the perturbed iron-opacity peak region, which also moves spatially inward in the simulation because of the readjusting temperature and density scales. Simultaneously, a self-consistent line-driven wind outflow is initiated from the regions slightly above the optical photosphere at around $\langle R_{sonic} \rangle$. Since the stellar flux is not constant in the wind initiation region, this outflow also starts to develop structure, leading to a complex interplay between structure formation deep in the atmosphere and in the overlying wind. After about $\sim 10 \ t_{\rm d,w}$ (or, $\sim$ 90 ksec) the simulation starts to reach a more dynamically relaxed state\footnote{By which we mean the simulations have relaxed in a statistical and global sense; the atmospheres remain time-dependent and dynamically very active throughout the simulations.}. \oc{As mentioned in the previous section, thermal relaxation of the deeper atmospheric layers may occur on different characteristic time scales. We examine this by taking lateral time-averages of the total model luminosity $\langle L_{\rm tot} \rangle_L$ along the radius, where $L_{\rm tot}$ is obtained from the stationary radial energy equation including full contributions from diffusive radiation, radiative and gas enthalpy (convection), as well as kinetic and gravitational energies (see Eq. (\ref{eq:ltot}) and the corresponding discussion below). If time-averages of $\langle L_{\rm tot} \rangle$ are sufficiently constant over radius, we may consider the simulation to be energetically relaxed (see also \citealt{goldberg_2022}). The colored curves in Fig. \ref{flux_variation} show laterally averaged total luminosities in the O4 simulation, for a time-average of 4 snapshots within $\sim t_{\rm d,w}$. By contrast, the black solid line shows a much longer average over the whole time range for which the colored lines were constructed, specifically for 
$\sim 100 t_{\rm d,w}$. As can be seen in the plot, the long time-average shows a fairly constant luminosity in radius, whereas the shorter averages display variations on the order of a few tens percent. Overall, this suggests that, in a statistical sense, our models are also energetically well relaxed.}   

After the initial adjustment period, Fig. \ref{time_series_relax} next displays time-snapshots from more advanced and dynamically relaxed stages of the O4 simulation. This shows how the $\Gamma < 1$ sub-Eddington layers close to the lower boundary have become quasi-stable, characterised by low variation in temperature and velocity; the radial velocity dispersion is here about $\sim 1 \rm \ km/s$ (see Fig. \ref{dispersion}) and the temperature dispersion is $\sim 1.5 \%$. As the temperature then decreases outwards from about 450 kK at the lower boundary, the opacity increases and creates local super-Eddington regions (below the photosphere) with $\Gamma > 1$. In accordance with previous simulations of this opacity-driven unstable zone \citep{Jiang_2015}, the net result is a highly structured and turbulent sub-surface atmosphere characterised by temperature fluctuations, strong density contrasts, and large turbulent velocities that \q{overshoot} into the overlying cooler surface layers with lower opacities. In contrast to previous work, however, in the simulations presented here the turbulent gas stemming from the deeper atmosphere naturally interacts with the line-driven layers around and above the photosphere. This complex interplay creates a situation wherein the line-driven wind acts like a suction mechanism, preventing some gas parcels that otherwise would have turned over from falling back into the deeper stellar envelope. Simultaneously, the turbulence arising from the deeper atmosphere introduces a natural variation in flux and line-opacity in the wind launching region. This leads to further structure formation and prevents the development of a stationary line-driven outflow as seen in 1D O star simulations with a similar Sobolev-based $\kappa^{\rm line}$ as here but assuming a static stellar surface and constant stellar luminosity \citep{luka_2022}.  The overall trend in behaviour is quite similar in all three models considered in this work. 

\begin{figure}
 \centering
 \includegraphics[width=9cm]{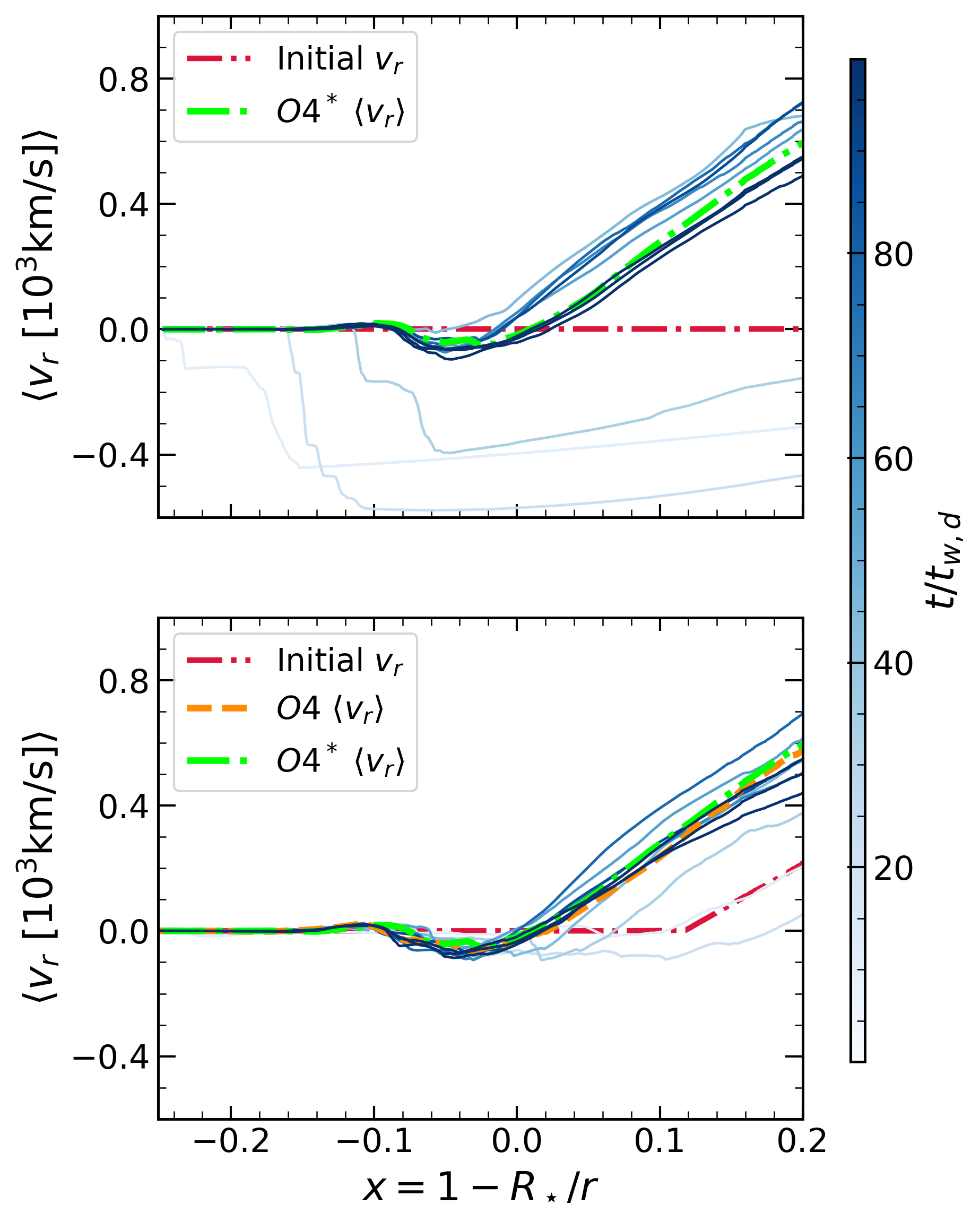}
      \caption{\oc{Running averages of the radial velocities for the O$4^{*}$ model (top) and O4 model (bottom), taken over 10 snapshots starting from the beginning of the simulation, with lighter colours representing earlier times. The red dash-dotted lines are the initial velocity profiles. The green dash-dotted line is the final average for the $\rm O4^*$ model and the orange dashed line is for the O4 model. See text.}}
    \label{vrun_avg}
\end{figure}

To examine how sensitive our models are to initial conditions, we re-run the O4 simulation but starting now instead from an essentially hydrostatic envelope (achieved in practice by assuming a negligible mass-loss rate $\sim 10^{-10} \ \rm M_\odot/yr$ and $\varv_\infty \sim 10 \ \rm km/s$ as boundary conditions in the 1D procedure outlined in the Sect. \ref{Initial_conditions}); that is, we start from a hydrostatic initial condition with effectively no wind to investigate if a line-driven outflow is still launched. Fig. \ref{vrun_avg} demonstrates the evolution of suitable running averages for radial velocity as a function of time, taken specifically for 10 snapshots over a few $t_{w,d}$ starting from the beginning of the simulation. Additionally, the figure also displays the initial conditions and a longer time-average taken after the wind has dynamically relaxed. The top panel shows the new \q{O$4^{*}$} model starting from the effectively no-wind condition (note the near-zero velocity curve for the initial condition); the bottom panel displays our standard O4 model. As discussed elsewhere in the paper, for this O4 model the adjustment of the atmosphere from the initial conditions involves inward movement of the photosphere. This leads to a similar adjustment of the line-driven wind, so that while the final average wind outflow is initiated from approximately the same optical depth layers as the initial condition this now occurs closer to the lower boundary radius. Without a significant initial velocity gradient, however, line-driving is at first quite inefficient in the O$4^{*}$ simulation and the whole upper atmosphere begins to fall inwards. Since this then gives rise to velocity gradients, line-driving starts to kick in and an outflow is subsequently developed during later snapshots. Comparing the final wind velocity curves of the O$4^{*}$ and O4 models, the only significant difference in the end is that the O$4^{*}$ model, as expected, takes much longer to develop a statistically relaxed outflow. This shows that the resulting line-driven winds in our O star simulations are quite robust against specific details in the initial wind conditions. 

\subsection{Model ${\rm O}4$} 
\label{model_o4}

\begin{figure*}
 \centering
 \includegraphics[width=18cm]{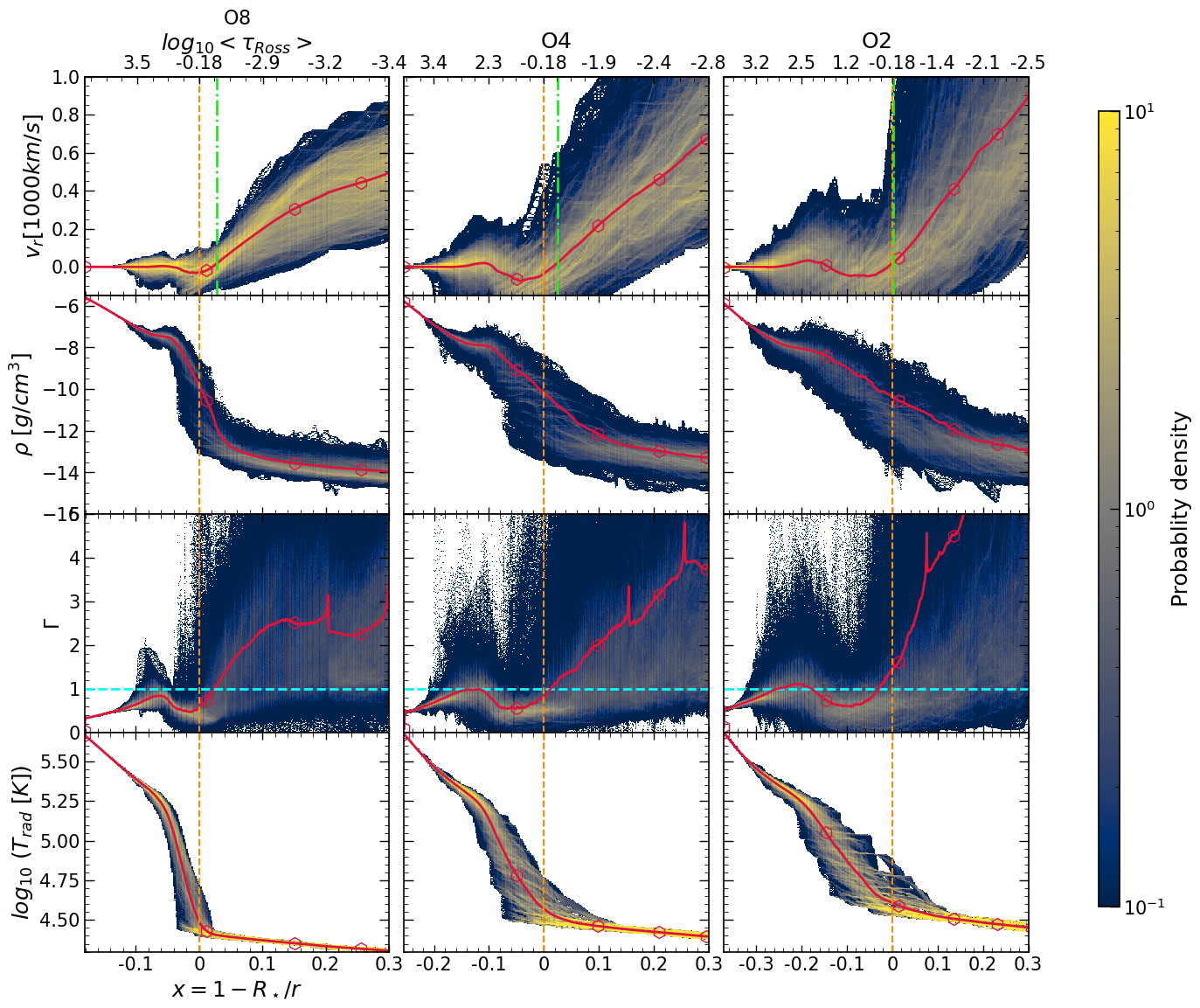}
      \caption{Probability density maps for different quantities, from top to bottom radial velocity, gas density, Eddington $\Gamma$, and radiation temperature for different models O8 (left), O4 (middle), and O2 (right). 40 snapshots have been used to create these probability density maps. The red curves represent the lateral-temporal averages. The vertical orange-dashed line is $\langle R_\star \rangle$, whereas the green dash-dotted vertical lines in the top panel are $\langle R_{sonic} \rangle$. The cyan horizontal lines in the $\Gamma$ panels are where its value is unity. The colours signify the probability of finding a quantity at a certain cell, with yellow having a higher probability and blue having a lower one. }
    \label{prob_grid}
\end{figure*}

We go on to examine in greater detail our prototypical model O4 for an early O-supergiant in the Galaxy. As discussed above,  after the initial relaxation, the O$4$ simulation instigates a turbulent deeper atmosphere driven by the iron-opacity peak, with line-driving taking over slightly above the optical photosphere to initiate an overall outflow. That is, although many gas parcels do indeed reach $\Gamma \ga 1$ already around the sub-surface iron-opacity peak, the O stars considered in this paper are not able to drive a net radial wind outflow from these regions. This is a key difference between the O star models of this paper and the optically thick winds arising from the simulations of WR-stars in \citet{nico_2022b}.

In Fig. \ref{prob_grid}, we illustrate probability density profiles (analogous to normalised 2D histograms) for radial velocity, radiation temperature, gas density, and $\Gamma$ for the O$4$ model (middle panel). The colour coding for the plots represents the likelihood of finding a cell with a specific value for the given quantity ($\varv_r$ , $\rho$, etc.) at a given radial coordinate. Thereby, at any radius, values that are coloured yellow are more likely to appear in the simulation than values that are coloured blue. The over-plotted red curve is then the laterally and time  \q{averaged} quantity. As evident from Fig. \ref{prob_grid}, at the iron-opacity peak $\langle \Gamma \rangle$ indeed grazes unity before again decreasing as we move outwards in the atmosphere toward $R_\star$. Since there is no net-outflow from these sub-surface layers, this in turn produces a very shallow $\langle \rho \rangle$ profile in the region where $\langle \Gamma \rangle \sim 1$, however in contrast to the initial conditions (see Fig. \ref{perturbed_density}) there is no sharp inversion present but rather a plateau-like feature. As $\langle \Gamma \rangle$ then decreases, $\langle \rho \rangle$ starts to decrease more rapidly again until line-driving eventually becomes efficient enough to produce higher $\langle \Gamma \rangle$ (recall that essentially $\kappa^{\rm line} \sim \rho^{-\alpha}$, Eqs. \ref{Eq:sobolev}, \ref{Eq:kappaline}). When this occurs, $\langle \Gamma \rangle$ quickly reaches values well above unity, thereby launching a line-driven supersonic outflow. This general behaviour is then also reflected in the $\langle \varv_r \rangle$ profile. Namely, around the iron-opacity peak $\langle \varv_r \rangle$ first increases slightly to around $\sim 20 \rm \ km/s$. The following decrease in $\langle \Gamma \rangle$ then produces a region with a net-infall of material before line-driving takes over above the average photosphere. As such, the net effect is that at $R_\star$ we interestingly enough observe a negative $\langle \varv_r \rangle \sim - 40 \rm \ km/s$ in our simulation, indicating that the average O star surface is actually slightly infalling. 

This general situation further produces large velocity dispersions, as illustrated in Fig. \ref{dispersion}. Focusing again on the O4 model, we see how the radial velocity dispersion $\langle \varv_{\rm disp} \rangle$ indeed is close to zero in the deepest layers of our simulations but then increases rapidly once instabilities connected to the iron-opacity peak become effective. The moving gas particles in this region then shoot into the cooler upper atmosphere, interacting with the line-driven outflow launched from the turbulent surface layers. For the O4-model, $\langle \varv_{\rm disp} \rangle \sim 80-90 \ \rm km/s$ around $R_\star$, after which it increases even further in the wind. As further discussed below, this general behaviour is overall in good agreement with O star observations of photospheric optical absorption lines as well as UV resonance lines formed in the wind \citep{hawcroft_2021}. Let us also point out here that although the characteristic (sub-)photospheric $\langle \varv_{\rm disp} \rangle$ values in our simulations are above the \q{gas} sound speed, $c_{s,gas}$, they are still below the \q{total} sound speed, $c_s$, of the radiating atmosphere (see Eqs. (\ref{Eq:totalsound}) to (\ref{Eq:radsound})).

\begin{figure}
 \centering
 \includegraphics[width=9cm]{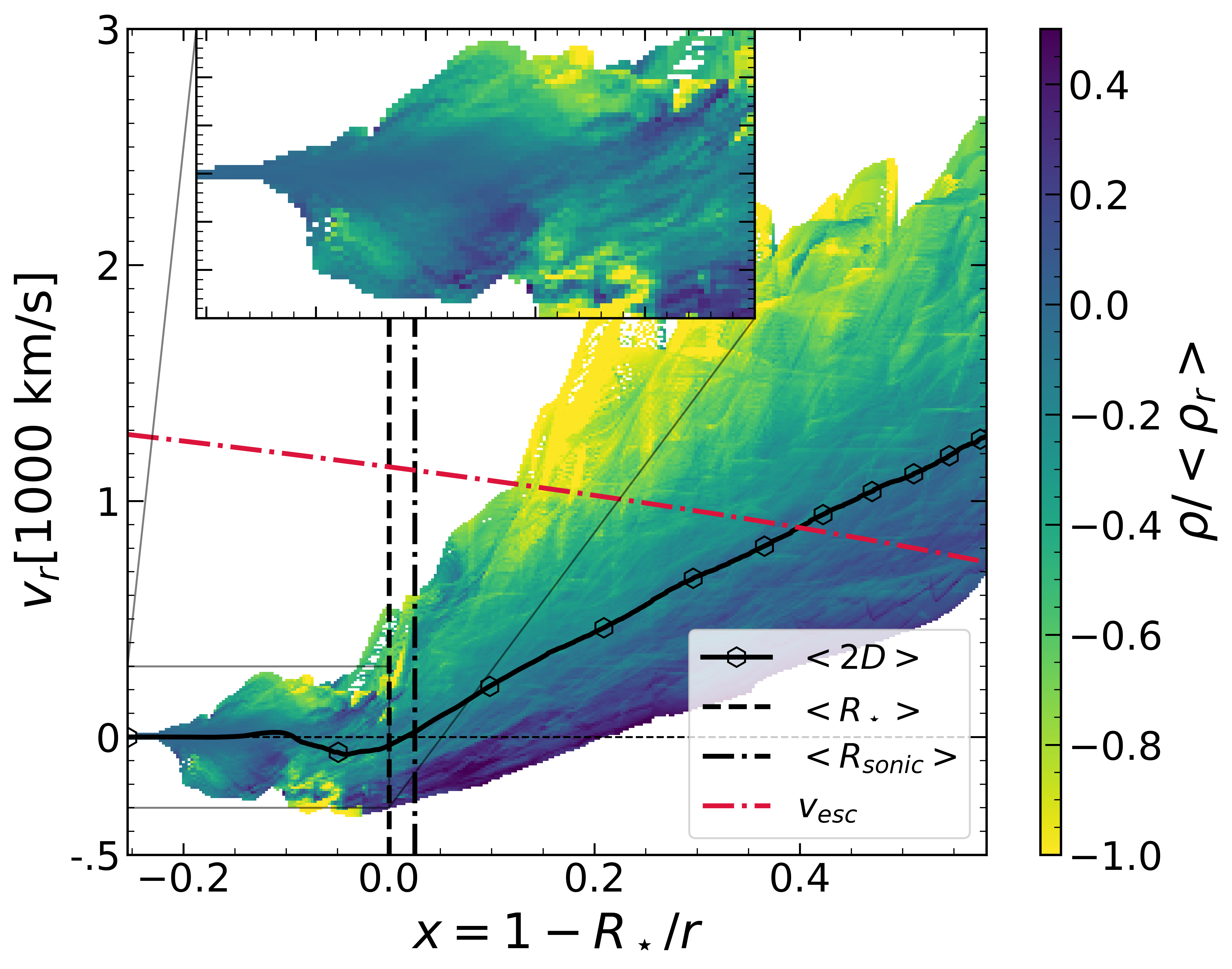}
      \caption{ Average relative density for each radial velocity at different radii for the O4 model. $\langle\rho_r\rangle$ is the average density at every radial cell. Colours here represent relative density ($\rho/\langle\rho_r\rangle$) with yellow displaying under-dense regions and blue representing over-dense clumps. The red dash-dotted line indicates the local escape velocity, the black dashed vertical line is the average photosphere $\langle R_\star \rangle$ and the black dash-dotted vertical line is the average sonic point $\langle R_{sonic} \rangle$. The zoomed-in part highlights the regions below the optical photosphere.}
    \label{turbulent_atmosphere}
\end{figure}

Figure \ref{turbulent_atmosphere} illustrates the average relative density for each radial velocity at different radii for the O4 mode. Blue here represents high-density clumps and yellow represents the under-dense structures. Although the under-dense structures reach their local escape velocities very fast and close to $R_\star$, on average the wind only escapes much further out. This is because the under-dense structures move at much higher velocities than their dense counterparts. As discussed above, line-driving is the main driver of the wind outflow here. However, since line-opacity and gas density essentially are inversely proportional to each other (see above), under-dense structures experience much stronger line-driving than dense clumps, resulting in significantly higher velocities for the former. The upshot is a strong anti-correlation between density and velocity in the O star winds studied here; that is, dense clumps in the wind move significantly slower than the rarefied gas in between them (see also \citealt{nico_2022b} for a similar result for WR-stars). By contrast, below $R_\star$ the gas is quite dense, line-driving is not efficient, and $\kappa$ thus does not have this direct inverse dependence on density. This then disrupts the above mechanism, resulting in a turbulent atmosphere without any such clear (anti-) correlation between velocity and density.

\subsection{Models O2 and O8} 
\label{model_o2_o8}

Here, we discuss two additional models, one (O2) with a higher luminosity and thus higher classical $\Gamma_e = \langle L_\star \rangle/L_{\rm edd}$, resulting in a higher $\langle T_{\rm eff} \rangle$, and one (O8) with a lower mass and luminosity, thus resulting in a lower $\Gamma_e$ and $\langle T_{\rm eff} \rangle$.  We computed these models by means of the same procedure; however, by varying the input parameters of the initial condition calculations, we can obtain different combinations of (lower boundary and stellar) radius, luminosity, and so on, for the simulations; the resulting $\langle 2D \rangle$ fundamental parameters of these models are provided in Table \ref{table:Models}.  

Inspecting once again the probability density clouds of Fig. \ref{prob_grid}, we observe that all three models exhibit similar qualitative behaviour. The lower boundaries for all models are (quasi-)stable, significant structures start developing around the iron-opacity peak, and line-driven winds are initiated around and slightly above the stellar surface. However, due to its lower baseline classical Eddington ratio, $\langle \Gamma \rangle$ of the cooler O8 simulation on average stays slightly below unity throughout the iron-opacity peak opacity region, leading to somewhat lower (but still significant) dispersion in the atmosphere. Vice versa, the hotter O2 simulation has a higher Eddington ratio, and as such $\langle \Gamma \rangle$ again grazes (even slightly exceeds) unity already in the sub-surface layers. But since the atmosphere there still is too dense for efficient line-driving, $\langle \Gamma \rangle$ drops again once the iron-opacity peak is overcome so that a deep sub-surface net-outflow is not launched, mimicking the O4 model. We note, however, that for this very luminous model, it is somewhat \q{easier} for line-driving to become efficient, meaning a lower line-opacity force-multiplier boost is needed to overcome the effective gravity and launch an outflow. As such, in the O2 simulation, the average sonic point $\langle R_{\rm sonic} \rangle$ lies at $\langle \tau_{\rm Ross} \rangle = 0.62$, meaning the average wind is on the verge of becoming (marginally) optically thick, and also that the characteristic region with negative average velocity now is located slightly beneath the stellar surface. Again this demonstrates how the simulations presented here, and previously in \citet{nico_2022b}, are able to very naturally capture the transition from lower luminosity stars with optically thin line-driven winds (O stars, hot sub-dwarfs) to higher-luminosity objects with optically thick winds ( \q{slash} or WNh-stars, classical WR-stars).

\begin{figure}
 \centering
 \includegraphics[width=9cm]{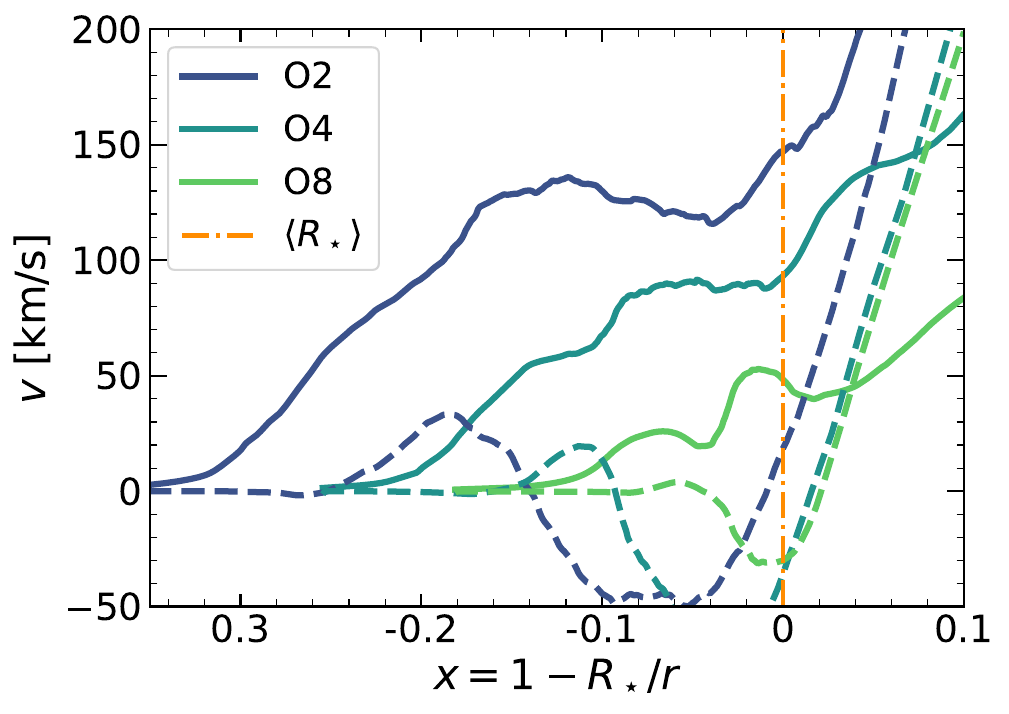}
      \caption{2D average radial velocity (dashed lines) and its dispersion (solid lines) as a function of the scaled radius $x= 1 - R_\star/r$. The colours correspond to the different models.}
    \label{dispersion}
\end{figure}

The variability in the sub-surface layers of the three simulations also follows simple qualitative trends with increasing classical Eddington ratio, $\Gamma_e$; the dispersion in density, temperature, and velocity is lowest for the O8 simulation and highest for the O2 model. Specifically, Fig. \ref{dispersion} illustrates how $\langle \varv_{\rm disp} \rangle$ around the photosphere exceeds 100 km/s in the O2 model, is below 50 km/s in the O8 model, and (as discussed already above) lies in between for the O4 model.

\section{Comparison to 1D models}
\label{1d_comparison}

\begin{figure}
 \centering
 \includegraphics[width=9cm]{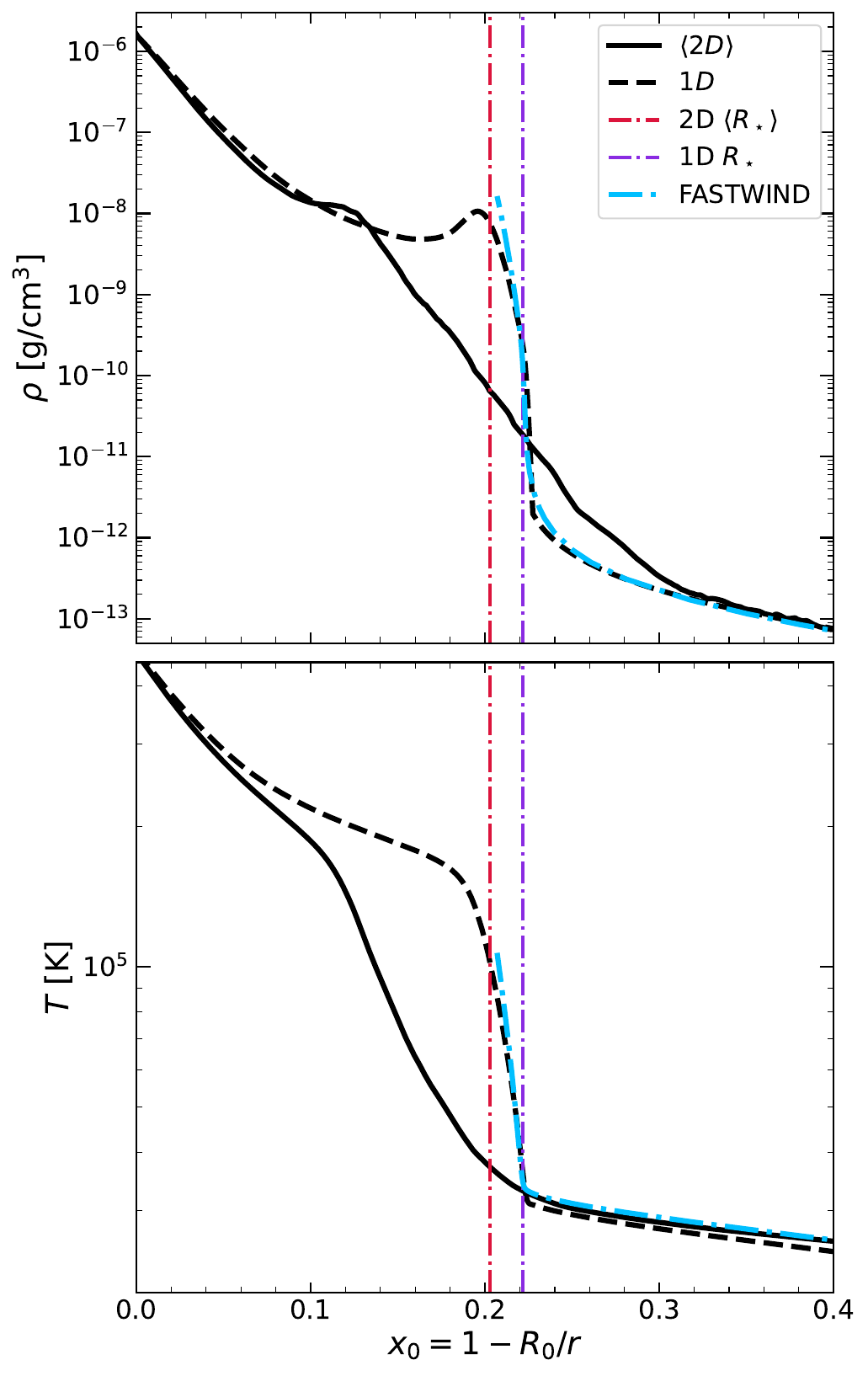}
      \caption{Initial 1D and resulting $\langle 2D \rangle$ gas density [in $\rm g/cm^3$] (top) and temperature [in K] (bottom) structure for the O4 model as a function of modified radius $x = 1- R_0/r$, with $R_0$ as the lower boundary radius. The \q{red dash-dotted} line is the $\langle 2D \rangle$ photosphere and the \q{blue dash-dotted} is the 1D photosphere. The cyan dash-dotted curve is produced using {\fontfamily{qcr}\selectfont FASTWIND} as described with specifications in Sect. \ref{Initial_conditions}.}
         \label{init_density_temp}
\end{figure}

\begin{figure}
 \centering
 \includegraphics[width=9cm]{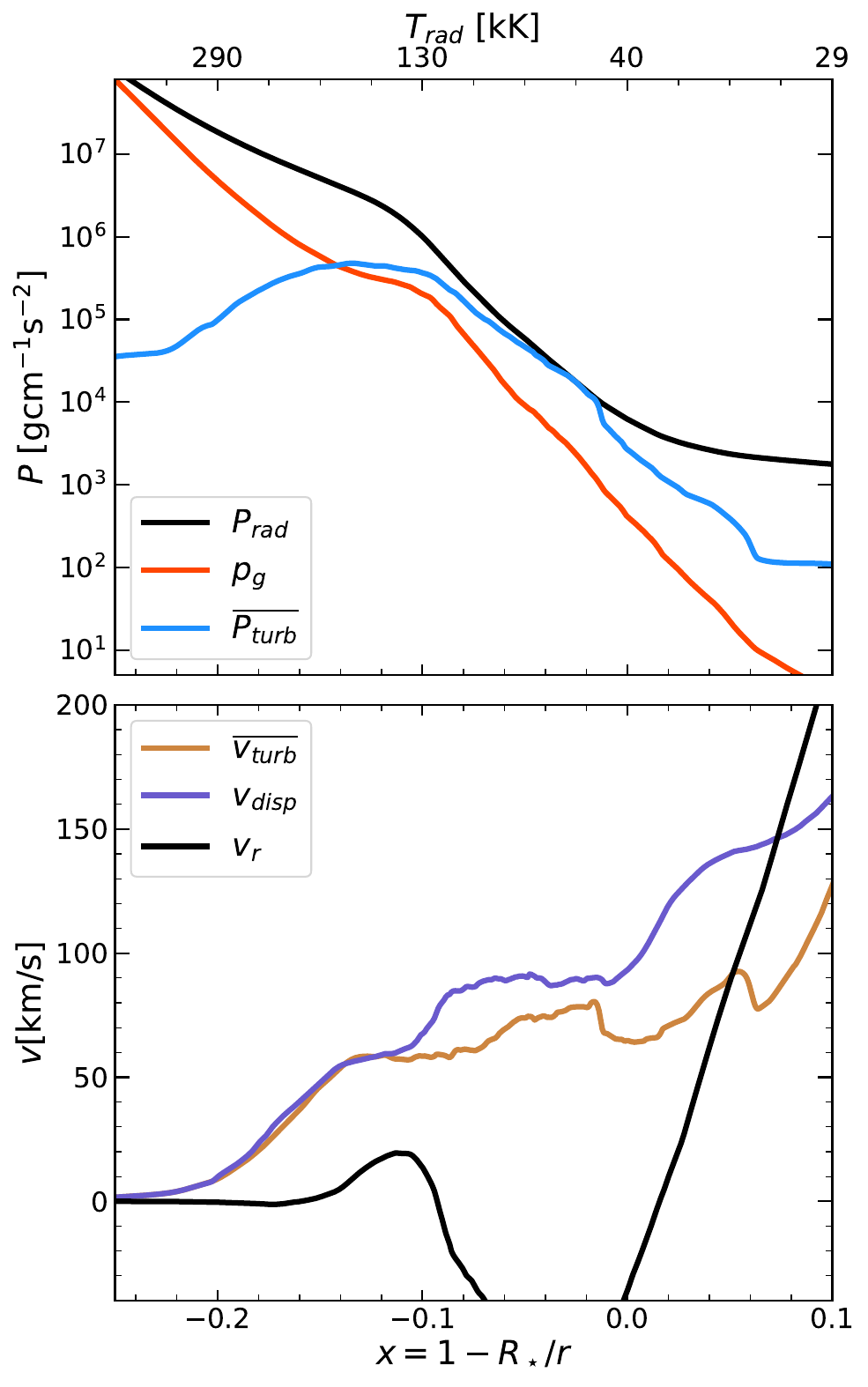}
      \caption{2D average gas pressure, radiation pressure, and turbulent pressure (top panel) and turbulent velocity, dispersion of radial velocity, and radial velocity (bottom panel) for the O4 model as a function of the scaled radius $x= 1 - R_\star/r$ with the upper axis as radiation temperature in $10^3 \rm K$.}
    \label{pressure}
\end{figure}

The previous section demonstrated the development of a very turbulent, time-dependent O star atmosphere coupled with a wind outflow. We go on to examine how the \q{average} properties of these atmospheres may compare to standard 1D, stationary models. 

Fig. \ref{init_density_temp} compares the $\langle 2D \rangle$ density and temperature structure of the O4 model to those of 1D models computed by means of the procedures outlined in Sect. \ref{Initial_conditions}. We note that (for visualisation purposes) the abscissa in this figure displays a scaled radial coordinate $x_0$ and that the $\langle 2D \rangle$ and 1D models here have been calibrated such that they have the same $R_0$. As we do not know in advance the emergent average parameters of the stellar wind, such as mass loss rates and terminal wind speeds, the outer part of the 1D initial conditions will not generally correspond to our 2D model. Hence, we re-ran the 1D model with the same stellar parameters ($L_\star$, $M_\star$, and $R_0$) but with the updated emergent wind properties, thus allowing us to make a fair comparison. The figure clearly shows how the $\langle 2D \rangle$ model has significantly less envelope expansion than the 1D model, and thereby also a lower photospheric radius $R_\star$ and higher $T_{\rm eff}$. Specifically, for the $\langle 2D \rangle$ and 1D O4 models we obtain $R_\star = 16.7 R_\odot$ and $18.7 R_\odot$; $T_{\rm eff} = 39.6 \rm kK$ and $37.2 \rm kK$, respectively. 
Moreover, the sharp density inversion seen in the 1D initial conditions is no longer present in the $\langle 2D \rangle$ density structure, and the slope of the density profile is significantly shallower in the $\langle 2D \rangle$ model around the photosphere (i.e. the characteristic photospheric scale height is larger). This indeed suggests that processes typically not included in standard 1D unified model atmosphere with wind codes may be important for also the average structure of these key atmospheric layers.  

\subsection{Convective energy transport and turbulent pressure}
\label{convective_energy}

\paragraph{Modified 1D models.} Spurred by the findings above and based on an interpretation of the structures observed in the deep atmosphere of our 2D simulations as primarily a result of turbulence due to the convective instability \citep[see also discussion in ][]{Jiang_2015}, we have incorporated a simple treatment of convection and turbulence in our 1D code for computing initial conditions, following \citealp[][their Chapter 7]{kip_2013}\footnote{In practice, our implementation of this formalism follows a \q{koch rezept} stemming from a set of notes developed by Jo Puls.}. This treats energy transport within the standard mixing-length theory (MLT) framework for non-adiabatic convection, accounting for the effects of radiation pressure and cooling. The free input parameter, $\alpha_{\rm MLT} \equiv \ell/H_{\rm p}$, is the standard MLT parameter for mixing length, $\ell$, and \q{total} pressure scale height, $H_{\rm p} = |P/(dP/dr)|$, and, for simplicity, we have applied the standard Schwarzchild criterion for when convective energy transport occurs (rather than the more rigorous criterion discussed in Sect. \ref{perturbations}). Additionally, we have also included a turbulent pressure term $P_{\rm turb} = \rho \varv_{\rm turb}^2$ in the hydrostatic Eq. (\ref{Eq:HE}), where $\varv_{\rm turb}$ is an input turbulent velocity parameter, such that now $dP/dm = GM_\ast/r^2$ for total pressure $P = p_{\rm g} + P_{\rm rad} + P_{\rm turb}$. The connection of the sub-surface layers to the overlying analytic wind outflow then follows the same procedure as before.

To make suitable comparisons, we first calibrated the new 1D models such that their $T_{\rm eff}$ and $R_\star$ now agree with the corresponding $\langle 2D \rangle$ simulations. If this calibration is done without the addition of convective energy and turbulent momentum transport, the result is again that the 1D models have a much more inflated envelope, a density inversion around the iron-opacity peak, and a steeper density-profile around the photosphere.

First adding convective energy transport by setting $\alpha_{MLT} = 1.0$, but still keeping $\varv_{\rm turb}= 0 \ \rm km/s$, gets rid of the density-inversion also in the 1D structure. However, around the photosphere (above the convectively unstable region) the scale-height is still too small, producing the same steep density-profile. 

\begin{figure}
 \centering
 \includegraphics[width=9cm]{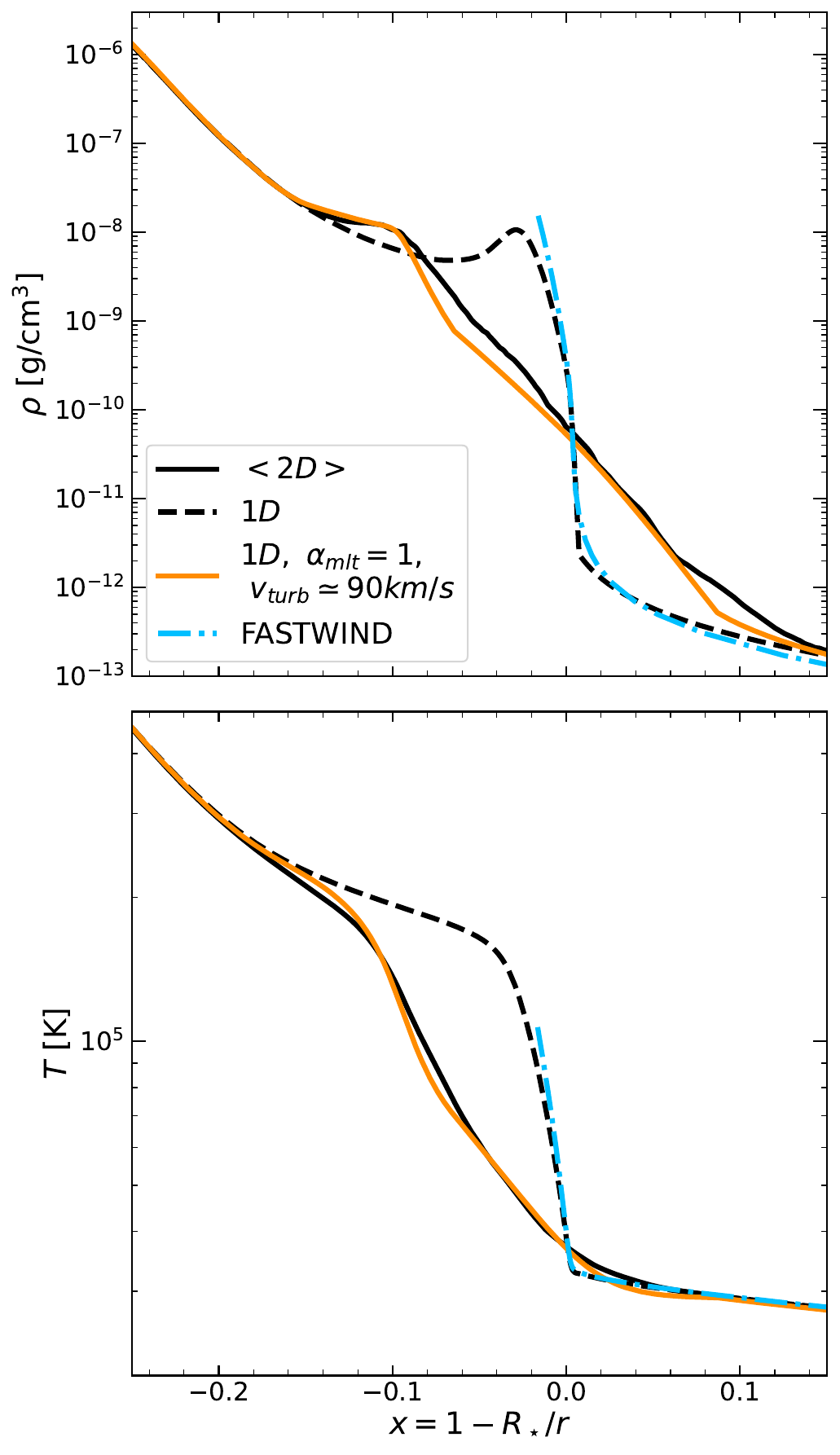}
      \caption{$\langle 2D \rangle$ gas density [in $\rm g/cm^3$] (top panel) and temperature [in $K$] (bottom panel) as a function of the scaled radius $x = 1- R_\star/r$ for the O4 model. The 1D structure is calibrated to the $\langle T_{eff} \rangle$ and $\langle R_\star \rangle$ obtained from the 2D model. The orange line displays the calibrated 1D model with an $\alpha_{MLT} = 1 $ and a turbulent velocity of $90 \ \rm km/s$ at the photosphere and reduced gradually in the deeper regions close to the hydrostatic envelopes. The cyan dash-dotted curve is produced using {\fontfamily{qcr}\selectfont FASTWIND}, using the specifications outlined in Table \ref{table:Models} and Sect. \ref{Initial_conditions}.}
  \label{density_temperature_comparison}
\end{figure}

In the next step, we first add a constant $\varv_{\rm turb} \simeq 90$ km/s throughout the atmosphere in our 1D model. This then indeed produces a shallower density-slope around the photosphere, but due to the constant turbulent velocity throughout the complete atmosphere, we (re-)introduce the envelope expansion of the lowermost layers. Finally, then, we introduce a simple variation of $\varv_{\rm turb}$ such that at the photosphere we have a maximum $\varv_{\rm turb} \simeq 90$ km/s which then gradually decreases below the photosphere to essentially 0km/s in the lowermost layers. This produces a reasonable agreement between the 1D and $\langle 2D \rangle$ results, as illustrated in Fig. \ref{density_temperature_comparison}. 
It is important to here note that we have not aimed for a perfect match of the $\langle 2D \rangle$ and 1D structures. Rather, our goal has merely been to demonstrate that the overall effects seen in the sub-surface layers of the 2D simulations may, in principle, be captured by these modifications of present-day 1D models. More quantitative comparisons and benchmarks will, however, require more in-depth analysis of the detailed structures seen in a larger array of simulations. Moreover, we have here not touched upon how the rather complicated atmospheric $\langle 2D \rangle$ average velocity field might be captured in such 1D stationary models (e.g. the negative average radial velocity around the photosphere and the subsequent rather complex wind initiation region). 

Nonetheless, it is quite striking how much steeper the density and temperature slopes around the photosphere are in the {\fontfamily{qcr}\selectfont FASTWIND}-like 1D comparison models. We may understand this large difference by comparing the characteristic density scale height in a model accounting for turbulent pressure to that in a purely radiative model, $H_\rho^{\rm turb} \approx H_\rho^{\rm rad} (1 + \varv_{\rm turb}^2/a_{\rm g,iso}^2)$. For $\varv_{\rm turb} \sim 90$ km/s and $a_{\rm g,iso} \sim 30$ km/s (appropriate for the photospheric layers) then, we get thus $H_\rho^{\rm turb} \approx 10 H_\rho^{\rm rad}$. That is to say, the density scale height is increased by a full order of magnitude, explaining the significantly different slopes seen in the models displayed in Figs. \ref{init_density_temp} and \ref{density_temperature_comparison}.  Moreover, since the turbulent velocities in our 2D simulations decrease rapidly below the iron opacity peak, the effective density scale height there becomes significantly reduced, leading to less envelope expansion than in 1D comparison models with high $\varv_{\rm turb}$ also in the lowermost atmosphere. As further discussed in Sect. \ref{discussion}, these differences may have a significant effect on spectral line formation and thus spectroscopic determination of fundamental parameters in the O star regime. \\

\begin{figure}
 \centering
 \includegraphics[width=9cm]{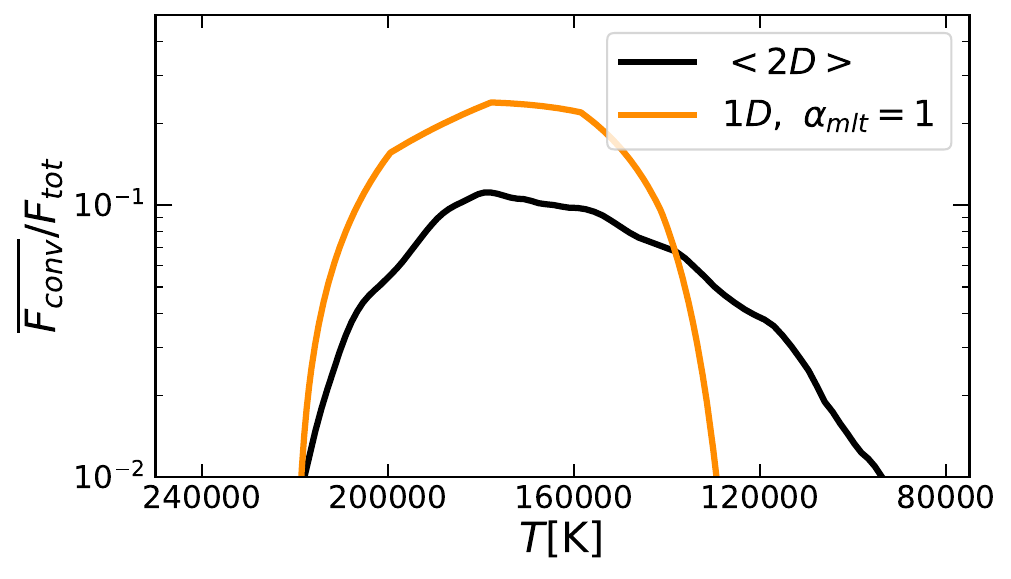}
      \caption{Convective flux to the total flux as a function of temperature for the O4 model. The black curve is the $\langle 2D \rangle$, and the orange curve is the calibrated 1D model to match the 2D with $\alpha_{MLT} = 1 $.}
  \label{flux_quantities}
\end{figure}

\paragraph{2D model averages.} The parameters introduced and values used above to (at least qualitatively) reproduce the average density and temperature profiles of the $\langle 2D \rangle$ atmosphere may further be directly compared to and interpreted through the radiation-hydrodynamic equations. Taking the stationary, radial components of the combined gas and radiation momentum and energy equations give \oc{\citep{mihalas_1984, turner_2001, goldberg_2022, jiang_2023}}
\begin{equation}
    \frac{\partial}{\partial r} 
    \left( \rho \varv_r^2 + p_{\rm g} \right) = - \rho \frac{G M_\ast}{r^2} + \rho \frac{\kappa F_{\rm diff}}{c} , 
    \label{eq:ltot}
\end{equation}
\begin{equation} 
    \varv_r \left( p_{\rm g} + P_{\rm rad} 
    + e_{\rm g} + E_{\rm rad} + \frac{\rho \varv_r^2}{2} - \rho \frac{G M_{\ast}}{r}  \right) +F_{\rm diff}  = F_{\rm tot} = \frac{L_{\rm tot}}{4 \pi r^2} ,
\end{equation}
where $L_{\rm tot} = \rm const.$ is the total luminosity, and where we have assumed an isotropic velocity field in the momentum equation. We note further that for the case $\gamma_g = 5/3$ and $3 P_{\rm rad} = E_{\rm rad}$, which is a good assumption in the deep optically thick layers, the stationary energy equation above may be equivalently written as: 
\begin{equation} 
    \dot{M} 
    \left( h_{\rm g} + h_{\rm r} + \frac{\varv^2}{2}- \frac{G M_{\ast}}{r}\right) + L_{\rm diff} = L_{\rm tot}, 
\end{equation} 
for gas and radiation enthalpies $h_{\rm g} = (5/2)p_{\rm g}/\rho$ and $h_{\rm r} = 4 P_{\rm rad}/\rho$, respectively, and diffusive luminosity $L_{\rm diff} = 4 \pi r^2 F_{\rm diff}$ \citep[e.g. ][]{owocki_2017}. These stationary relations invite us to identify the following suitable lateral- and 
time-averages as corresponding to \q{turbulent pressure} and \q{convective flux} in our simulations: 
\begin{align}
    & \overline{P_{\rm turb}} = \langle \rho \varv_r^2 \rangle,  \\
    & \overline{F_{\rm conv}} = 
    \langle \varv_r(p_{\rm g} + P_{\rm rad} 
    + e_{\rm g} + E_{\rm rad}) \rangle ,
\end{align}
with associated turbulent and convective velocities 
\begin{align}
  &  \overline{\varv_{\rm turb}} = 
    \sqrt{ \frac{\overline{P_{\rm turb}}}{\langle \rho \rangle} } ,  \\
  &  \overline{\varv_{\rm conv}} = \frac{\overline{F_{\rm conv}}} 
    {\langle p_{\rm g} + P_{\rm rad} 
    + e_{\rm g} + E_{\rm rad} \rangle }.  
\end{align}
That is, the turbulent velocity is here simply identified as the density-weighted average root mean square (rms) velocity \citep[see also][]{goldberg_2022}. As mentioned above, this expression for $\overline{P_{\rm turb}}$ essentially assumes that the turbulent velocity field is isotropic in the relevant layers (specifically for our case that $\langle \rho \varv_r^2 \rangle \approx \langle \rho \varv_t^2 \rangle$), or alternatively (as pointed out in \citealt{jiang_2023}) that the scale height is much smaller than the radius; these conditions are generally true in the sub-surface layers of our simulations that do not experience a net radial outflow. On the other hand, the appropriate convective velocity is rather identified from the energy equation, and as such in general 
$\overline{\varv_{\rm turb}} \ne \overline{\varv_{\rm conv}}$. For the O star simulations here the additional kinetic and gravitational energy fluxes are small, and $\overline{F_{\rm conv}}$ is now essentially what MLT tries to estimate  (see also discussion in \citealt{jiang_2023}).
As can be seen in Fig. \ref{flux_quantities}, the maximum amount of energy transported by $\overline{F_{\rm conv}}$ in our O4 model is $\sim 10 \%$ of the total flux. It is dominated by the radiative enthalpy component, with gas enthalpy providing only $\sim 0.1 \%$ of the total flux. $\overline{F_{\rm conv}}$ is further centered around the iron-opacity peak regions, but with a soft upper boundary that reaches significantly cooler layers than the corresponding 1D model. For this 1D model with $\alpha_{\rm MLT} = 1$ (calibrated to approximately match the $\langle 2D \rangle$ density and temperature profiles) the maximum energy transported via $F_{\rm conv}$ is $\sim 20 \%$. We note that to obtain similar energy transport by the convective flux in the $\langle 2D \rangle$ and 1D O4 models one only needs  $\alpha_{\rm MLT} = 0.5$. However, this is not enough to wash away the density inversion in our 1D model. This difference will require more analysis in a future work, but this effect may stem from the simple fact that our 1D model applies a simple Schwarschild criterion without \q{overshooting} for the boundaries of the convective zone. As such, the region of convective transport is much narrower than in the $\langle 2D \rangle$ model (see Fig. \ref{flux_quantities}), which then may require somewhat more efficient transport to obtain a similar net effect. 

For the O2 and O8 models, we find a similar behaviour of $\overline{F_{\rm conv}}$, but with maximum levels of $\overline{F_{\rm conv}}$ that reach $\sim 11 \%$ and $\sim 4 \%$ of the total flux from 2D respectively. Similar to the O4 model, gas enthalpy is quite insignificant in comparison to radiation enthalpy for both models. 
As evident from Fig. \ref{pressure}, for the O4 model, in the layers leading up to the photosphere, radiation and turbulent pressures dominate the total pressure balance. As we approach the deeper layers of the stellar envelope, the turbulent pressure becomes lower so that we have a clear radiation-dominated atmosphere around the iron-opacity peak. As the opacity then decreases again, the turbulent pressure declines rapidly and the gas pressure, $p_{\rm g}$, starts to become somewhat more significant, approaching (though remaining below) the level of radiation pressure at the lower boundary. In the bottom panel of Fig. \ref{pressure}, we display the turbulent velocity $\overline{\varv_{\rm turb}}$ for the O4 model. The qualitative behaviour of $\overline{\varv_{\rm turb}}$ is quite similar to the straight average $\varv_{\rm disp}$ discussed in the previous section, with very similar values in the deep subsonic layers. Once we approach the upper atmosphere, we notice a difference in that the density-weighted $\overline{\varv_{\rm turb}}$ is somewhat lower than $\varv_{\rm disp}$. This is most likely related to the influence of the line-driven wind and the density-velocity anti-correlation discussed in the previous section. For the O4 model in the layers around the optical photosphere, we observe the turbulent velocities to be $\sim 60 - 80 \ \rm km/s$. The general behaviour seen in the simulation is, thus, quite consistent with the modification of the pressure balance introduced above in the corresponding 1D model. Interestingly, $\overline{\varv_{\rm conv}}$ only reaches a peak value (in the iron-opacity peak region) of $\sim 20 \ \rm km/s$, which is much lower than the turbulent velocity discussed above; effectively this means that the turbulent momentum transport in our simulations is much more efficient than the advective energy transport in the iron-opacity peak region. Moreover, below the optical photosphere, although the gas sound speed $c_{\rm s, gas}$ is around $\sim 30 \ \rm km/s$, the total sound speed $c_s$ is above $100 \ \rm km/s$. As such, both $\overline{\varv_{\rm turb}}$ and $\overline{\varv_{\rm conv}}$ stay well below the local total sound speed in our simulations. 

Broadly, we also observe the same trends for $\overline{\varv_{\rm turb}}$ in the O2 and O8 models as for the O4 simulation, with $\overline{\varv_{\rm turb}}$ at the photosphere  $\sim 100 \ \rm km/s$ for the O2 simulation and $\sim 30 \ \rm km/s$ for the O8 model. In the case of the O8 simulation, $p_{\rm g}$ and $P_{\rm rad}$ are quite similar in the deeper subsurface layers, with $p_{\rm g}$ even slightly above at the lower boundary. On the other hand, for the O2 the qualitative behaviour of $p_{\rm g}$ and $P_{\rm rad}$ are similar to the O4 model, although $\overline{P_{\rm turb}}$ approaches $P_{\rm rad}$ much quicker than in the O4 model. 

\begin{figure}
    \centering
    \includegraphics[width=9cm]{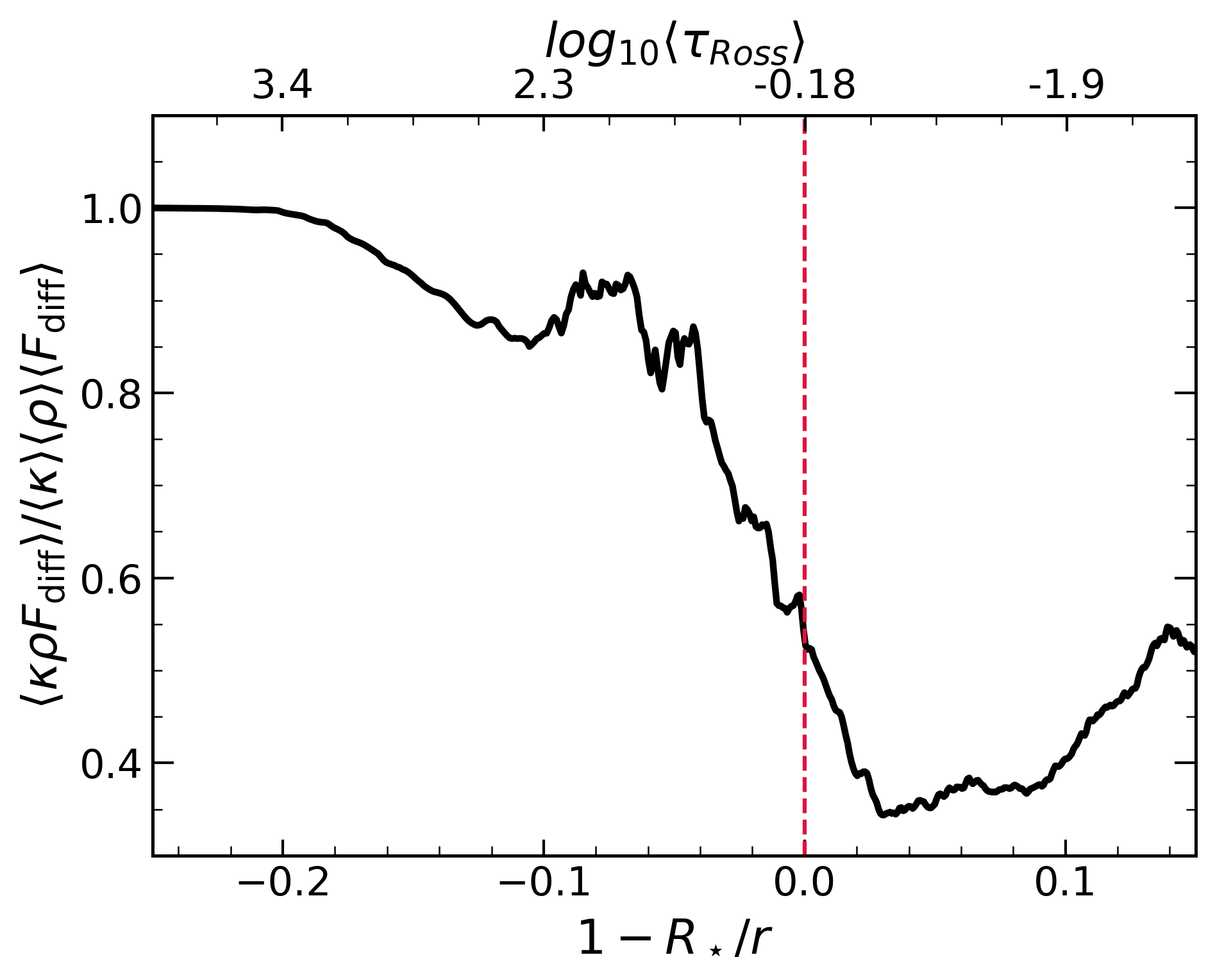}
    \caption{\oc{Flux-porosity effect as a function of scaled radius for our O4 model.}}
    \label{porosity}
\end{figure}

\subsection{Porosity reduction of radiation force}

Because of potential (anti-)correlations between density, radiative diffuse flux, and opacity, the averaged radiation force can be different in multi-dimensional simulations as compared to 1D models based on their individual averages. That is, generally 
$\langle \rho \kappa {\bf F_{\rm diff}} \rangle \ne \langle \rho \rangle \langle \kappa \rangle \langle {\bf F_{\rm diff}} \rangle$ (\citealt{shaviv_1998}, see also discussion in \citealt{jiang_2023}). Fig. \ref{porosity} displays this difference for our O4 simulation, illustrating that while the quasi-steady lowermost atmosphere shows no significant modification, the averaged radiative force becomes reduced in our 2D simulations when approaching the photosphere. This is in qualitative agreement with the analysis by \cite{schultz_2020} (based on the simulations by \citealt{Jiang_2015, jiang_2018}), although the reduction here is quantitatively somewhat lower. We have not included this effect in the 1D comparison models above, also because in layers where the reduction is most prominent we follow the customary approach for 1D unified model atmospheres used for spectroscopy and apply an analytic $\beta$-type velocity law (see previous sections). The effect could be significant also for deeper quasi-static layers, however, and thereby modify the overall pressure gradient balance discussed above; this should be investigated in detail in future work when trying to further calibrate 1D atmospheric models (see also discussion in \citealt{schultz_2020}). Moreover, we note that these types of \q{porosity effects} may be important for wind launching in other domains \citep{owocki_2004} as well as for capturing spectrum synthesis effects stemming from the clumpy atmosphere and wind within stationary 1D models (see discussion in Sect. \ref{wind_turbulence}). 

\subsection{Mass loss rates}
\label{mass_loss}

\begin{figure}
 \centering
 \includegraphics[width=9cm]{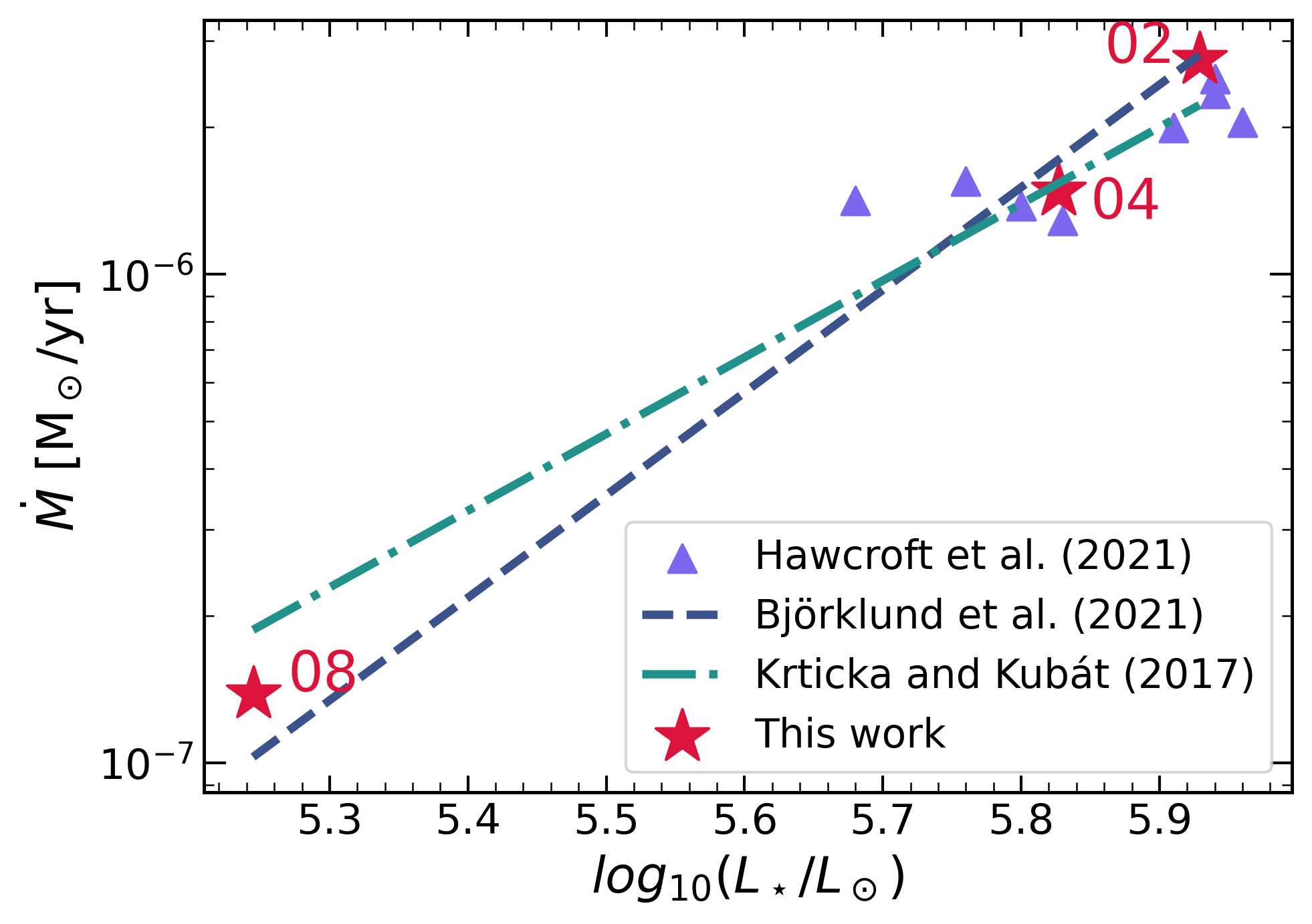}
      \caption{Mass loss rates in $\rm M_\odot/year$ as a function of solar luminosity. The $\star$-markers are the mass loss rates calculated from the models considered in this work, whereas the dot-dashed line using \cite{krticka_2017}, and the dashed line is calculated using \citet{robin_2021}. The triangle markers are plotted from the empirical mass loss rate study by \citealt{hawcroft_2021} (their optically thick GA rates).}
  \label{massloss_rates}
\end{figure}

Additionally, we compared the averaged mass loss rates 
$\langle \dot{M} \rangle$ (see Table \ref{table:Models}) of our models to mass loss rates derived by dedicated 1D, stationary theoretical modeling \citep[here][]{krticka_2017,robin_2021}. Since we are using chemical abundances prescribed by \citet{grevesse_1993} we applied a simple metallicity scaling 0.02/0.013 in the O star recipe by \citet{robin_2021}, to reflect the different baseline solar abundance scale. Similarly, for the \citet{krticka_2017} recipe, we also applied this metallicity-scaling based on \citet{robin_2021}.
Although this re-calibration is not perfect \citep[since changes in abundances of important driving ions do not necessarily follow the baseline metallicity scaling, see discussion in ][]{sundqvist_bjorklund_2019}, it nonetheless captures the principal effect. Additionally, we also compared to the recent empirical study of luminous Galactic O-supergiants by \citet{hawcroft_2021}. The overall good agreement between the average
$\langle \dot{M} \rangle$ computed here and the rates predicted by these (albeit 1D, stationary) more focused studies (see Fig. \ref{massloss_rates}) provide further support that we indeed are able to capture the line-driving effect rather well within our new modeling framework (see also \citealt{luka_2022}). 

In this context, it is further important to recall that comparisons of \q{predicted} $\langle \dot{M} \rangle$ values cannot typically be carried out against standard 1D unified model atmosphere codes used for spectroscopic studies; this is because in these cases, $\dot{M}$ is most often (though see, e.g. \citealt{sander_2017}) treated as a free input-parameter accompanied by a parameterised analytic velocity field (see details previous sections). By contrast, in the simulations presented here $\langle \dot{M} \rangle$ and the atmospheric velocity field are intrinsic and self-consistent emergent properties of the models. 

\section{Discussion} 
\label{discussion}

\subsection{Convective energy transport}
\label{conv_transport}

As shown in Sect. \ref{convective_energy}, radiation enthalpy is able to carry only a small part of the total flux through the sub-surface iron opacity peak region of our model O stars. This means our simulations are generally located in the region of so-called \q{inefficient} convection, which we may interpret through some simple scaling relations. For the radiation-dominated atmospheres studied here, an upper limit for convective flux would scale as \citep[e.g. ][]{grafener_2012} $F_{\rm conv} \sim c_s E_{\rm rad}$, so that $F_{\rm conv}/F_{\rm diff} \sim (c_s/c) (T/T_{\rm eff,0})^4 \sim (c_s/c) \tau$, where $T_{\rm eff,0}^4 = L_\star/(\sigma 4 \pi R_0^2)$ is an effective (flux) temperature evaluated below the iron-opacity peak at $\sim R_0$ and where the generic scaling $\tau \sim (T/T_{\rm eff,0})^4$ has been applied further. Alternatively, we may view this scaling through the characteristic dynamical and radiative diffusion timescales for the region, $t_{\rm d,a} \sim H_{\rm p}^{\rm tot}/c_s$ and $t_{\rm diff} \sim (\tau/c) H_{\rm p}^{\rm tot}$, such that $t_{\rm diff}/t_{\rm d,a} \sim (c_s/c) \tau$. That is, when the diffusion timescale is shorter than the dynamical timescale, gas particles quickly adjust to their surroundings such that energy transport by means of convection (advection) is no longer efficient for carrying a significant fraction of the total flux entering from the quasi-stable radiative zone below. Setting $t_{\rm diff} = t_{\rm d,a}$ then also gives us the \q{critical optical depth} $\tau_c = c/c_s$ as defined and used in \citet{Jiang_2015} to distinguish between efficient and non-efficient convective energy transport. Inspection gives $\tau$ on the order of hundreds for our simulations (see also corresponding figures displaying a mean optical depth scale), so that $F_{\rm conv}/F_{\rm diff} \sim t_{\rm diff}/t_{\rm d,a} \sim 0.1$ indeed is a quite good characteristic scaling number, as found in our O star simulations (see previous section). We note that for WR-stars, this number will likely be even lower owing to their higher core effective temperatures \citep{grafener_2012, nico_2022b} whereas for more extended and cooler LBV-like massive stars, the ratio of convective to diffusive fluxes may be significantly higher for the iron-opacity peak zone \citep{Jiang_2015, jiang_2018}.

Since the majority of the radiative flux thus cannot be carried by convective transport through the iron-opacity peak region, it means the radiative flux is only marginally lowered and that many gas particles there effectively have $\Gamma > 1$. As such, they can be accelerated to above the local gas sound speed, thereby producing the high values of turbulent velocities generally seen in our simulations. This is overall consistent with the results found by \citet{Jiang_2015}, whose simulation {\fontfamily{qcr}\selectfont StarTop} is the one that is closest to the parameters of our O star models here. For this simulation, also they find inefficient convective energy transport, with radiative enthalpy carrying less than a percent of the total flux. That number is even lower than found here, but likely this is largely caused by the different choices of the stellar input parameters defining the simulations. Moreover, similar to the results of this paper, they too find large turbulent velocities that exceed the local gas sound speed in their simulations.  

\subsection{Photospheric macroturbulence and microturbulence}
\label{micro_macro}

As discussed, in our models we find large atmospheric turbulent velocities, with the prototypical O4 model exhibiting $\sim 60 \ km/s$ around the optical photosphere; the O2 model has higher values than this, and the O8 simulation somewhat lower, thus marking a clear trend with $\Gamma_e$ (See Fig. \ref{dispersion}). This trend is in overall good qualitative agreement with observations of photospheric absorption lines of O stars, where it is typically necessary to apply a varying amount of \q{extra line-broadening} to match the line profiles computed by means of 1D atmospheric models to high-resolution observations. Interpreted through the (ad-hoc) fit parameter  \q{macroturbulence}, such studies do indeed find photospheric \q{macroturbulent velocities} that are on the same order as the turbulent velocities seen in our simulations ($\sim 30 -100 \, \rm km/s$), and moreover also find the same overall trend as here, that is higher macroturbulence for objects closer to the classical Eddington limit \citep[e.g. ][]{simon_2017}. 

Additionally, 1D models often apply a moderate amount of (also ad-hoc) photospheric \q{microturbulence} ($\sim 10-20 \, \rm km/s$ in standard {\fontfamily{qcr}\selectfont FASTWIND}-based analysis of O stars, e.g. \citealt{hawcroft_2021}), both when solving for the ionisation and excitation balance and when computing synthetic spectra. A distinction between micro- and macro-turbulence is usually motivated by means of optical depth arguments, as essentially borrowed from spectral analysis of cool, low-mass stars \citep[see detailed overview in][]{gray_2008}; in practice, this just means that microturbulence is explicitly included in the line-profile functions for computing number densities and synthetic spectra whereas macroturbulence is only applied afterward as a convolution that further broadens the spectral lines but preserves their equivalent widths \citep[thereby also causing degeneracy-problems between rotational and macroturbulent line-broadening, see][]{sundqvist_2013}. 

Overall, our simulations thus provide a natural explanation for the need for this ad-hoc \q{extra line-broadening} mechanism necessary to include in the 1D model atmosphere and spectrum synthesis codes \citep[which in a sense mirrors results found already for low-mass sun-like stars, e.g. ][]{asplund_2000}. It remains to be seen, however, how the large turbulent velocities observed in our simulations may (or may not) be captured in such standard 1D methods for performing quantitative spectroscopic analysis of O stars with winds. Certainly, quantitative comparisons would require spectral line synthesis calculations directly from our simulations, investigating not only the overall broadening of lines stemming from our models but also their predicted strengths, shapes, and variability due to complex dynamics. Some promising first results in this respect have been published by \citet{schultz_2022}, based on the simulations by
\citet{Jiang_2015, jiang_2018}, thus neglecting the influence of the wind outflow. 

\subsection{Wind turbulence and clumping}
\label{wind_turbulence}

The characteristic turbulent velocities in our simulations increase even further in the supersonic wind outflow. Again this qualitatively agrees with observations of UV resonance wind lines in O stars (so-called \q{P-Cygni} lines), which typically require turbulent velocities (usually included in the line-profile function as microturbulence, see above) that increase with wind velocity when fitted by means of 1D, stationary models, reaching terminal values on order $\sim (0.1-0.2) \varv_\infty$ \citep[e.g,][]{haser_1995, brands_2022, hawcroft_2023}.

Clumping factors $f_{\rm cl} \equiv \langle \rho^2 \rangle/\langle \rho \rangle^2$ range between $\sim 2-10$ in the outflowing wind parts of our simulations. While these are quite reasonable numbers, in this context it is important to recall that our method neglects the influence of the strong line-deshadowing instability \citep[LDI,][see further discussion in \ref{limitations}]{owocki_1984, owocki_1998, rybicki_1990}, which is believed to cause additional strong structure formation in O star winds. Moreover, for both Sobolev-based \citep{nico_2022b} and LDI-based 
\citep{sundqvist_2018, flo_2022} calculations it has been found that increasing the spatial dimensionality of the simulation typically decreases the quantitative values of $f_{\rm cl}$ in the wind. As such, we defer further more detailed quantifications of clumping factors in our simulations to future 3D modeling.  

Nevertheless, it is likely that the most important finding regarding clumping in our simulations concerns the \q{two-component medium} ansatz applied in (to our knowledge) all major spectroscopic 1D modeling attempts to account for wind clumping \citep{hillier_1991, oskinova_2007, sundqvist_2010, sander_2017, sundqvist_2018b}. Essentially, all such methods had previously relied on the assumption that the wind consists of small-scale density clumps embedded in a very rarefied (almost void) medium so that (almost) the whole wind mass is contained within the clumps (and the wind excitation and ionisation balance then can be derived only for the clumped medium). The recent studies by \citet{hawcroft_2021}, \citet{brands_2022}, and Verhamme et al. (in prep. for A\&A), however, all find strong empirical indications that the \q{interclump medium} seems to be much less rarefied than previously thought, and that a significant fraction of the total wind mass might not reside in overdense, small-scale clumps. Indeed, inspection of the probability colour maps in this paper really does not seem to find much basic support for a two-component medium ansatz; rather the most likely wind-density is close to the mean, with a large dispersion around it, and further with a strong anti-correlation between density and wind velocity. This was also found in the 3D WR models by \citet{nico_2022b} (see discussion therein) and similar distributions that are rather centered around the mean have actually also been observed in recent 2D LDI simulations \citep{flo_2022b}. 

While these preliminary notions will need further verification from more detailed and dedicated studies, in view of the above it is certainly the case that current assumptions underlying the treatment of wind clumping in present-day 1D model atmosphere codes are in need of a careful re-investigation. \oc{\citet{owocki_sund_2018} recently made a first such attempt to relax the standard two-component medium assumption, but further work is required since that study focused solely on continuum absorption, whereas the most important effects typically are observed for spectral line formation.}   

\subsection{General implication for 1D stellar atmosphere with wind models and spectral synthesis}
\label{implications_1d}

The most significant difference in the average photospheric structure of our 2D simulations and those of the 1D models are most probably related to the markedly different slopes of gas density and temperature around the optical photosphere (see Fig. \ref{density_temperature_comparison}). In our simulations, this effect scales with Eddington parameter $\Gamma_e$, with larger differences for the O2 and O4 models than for the less luminous O8 model. Additionally, the $\langle 2D \rangle$ models also display a significantly more complex average velocity structure around the photosphere and in the wind initiation region than usually assumed in the 1D models used for quantitative spectroscopy, as well as wind clumping properties that generally do not seem to support the basic assumptions normally adopted in 1D codes to account for such effects (see above). 

These differences will likely affect the formation process of essential spectral lines, thereby also affecting empirical spectroscopic derivation of fundamental stellar parameters ($T_{\rm eff}$, $g_\star$) as well as determinations of chemical abundances in the O star regime. Moreover, general calibrations and comparisons between evolution predictions and spectroscopic results may also be significantly affected, such as the so-called \q{mass-discrepancy problem} for O stars (e.g. \citealt{herrero_1992}). 

As shown previously, in order to reasonably match the $\langle 2D \rangle$ and 1D density and temperature structures, we need to in the latter add a large turbulent pressure $\overline{P_{\rm turb}}$ in the hydrostatic equation, and, moreover, a moderate $\overline{F_{\rm conv}}$ in the energy equation to get rid of density inversions around the iron opacity peak. It thus remains to be tested if these effects can be captured also without a downward extension of the lower boundary in current 1D O star atmospheric models such as {\fontfamily{qcr}\selectfont FASTWIND}, {\fontfamily{qcr}\selectfont PoWR}, {\fontfamily{qcr}\selectfont CMFGEN}, amongst others (which typically lies well above this region, see Sect. \ref{Initial_conditions}).

\subsection{Limitations of our simulations} 
\label{limitations}

To make these first multi-dimensional unified atmosphere and wind simulations for O stars possible, there are a number of simplifications we have made in our simulation techniques. Although we believe that none of them will drastically change the general nature of the results presented here, it will be an important focus of future work to try and improve upon some of them. 

First, as mentioned previously, we assume that the flux-, Plank-, and energy-weighted mean opacities are the same. In particular, this may affect $\dot{q}$, namely, the balance between radiative heating and cooling, since really only the pure absorption part of the total opacity should be accounted for (see also discussion in \citealt{nico_2022a}). While similar approximations (i.e. using the full $\kappa^{\rm line}$ for energy and Planck means) have been quite successfully applied to approximate effects of \q{line blanketing} in stationary 1D modeling of the temperature structure in optically thick WR winds \citep{lucy_1993}, it is likely that our current approach overestimates $\dot{q}$ in the wind-parts of the O stars studied here. Essentially, our present methodology introduces a strong wind blanketing effect that forces the radiation and gas temperatures to be almost identical; that is, our simulations effectively behave like LTE models. However, while this approximation has provided good estimates for the line force even in the O star regime \citep[e.g. ][]{luka_2022}, it is more questionable for computing $\dot{q}$ in such O star winds. As such, we have very recently attempted to extend our method for computing $\kappa^{\rm line}$ to (approximate) non-LTE conditions 
\citep[following][]{lucy_1985, lucy_1993, puls_2000}, separating the flux, energy, and Planck mean opacities. This method, and the first results regarding effects on the predicted energy balance and structure of our simulations, will be presented in an upcoming paper. 

Secondly, as already discussed in Sect. \ref{hybrid_opacity_sec} our hybrid opacity formalism may double-count line opacities in an \q{intermediate} region wherein the Sobolev-based $\kappa^{\rm line}$ has already started to become effective but $\kappa^{\rm Ross}$ still have contributions from line opacity. Since such situation may indeed occur around the regions where the wind is launched, it is natural to ask if this might significantly affect our predicted wind characteristics. In this respect, however, the fact that our average mass-loss rates overall agree well with those predicted by detailed 1D models based on full non-LTE and multi-frequency co-moving frame (CMF) radiative transfer (Fig. \ref{massloss_rates}) suggests that the impact is not severe, at least not in an average sense for the regime investigated in this paper. Specifically, as seen in Fig. \ref{massloss_rates} we obtain good agreement with the mass-loss rates predicted by \citet{robin_2021}. These models are based on the iterative method by \citet{sundqvist_bjorklund_2019}, using CMF solutions over the entire relevant frequency-range (typically between $\sim$  200 and 10\,000 $\AA$, see \citealt{puls_2020} for a detailed description) when computing the radiatvie acceleration, and also use the same line list as employed in this work. That is, if either our Sobolev-based line-driving calculations or the issues related to adding opacities had been severe, we would likely not find such good average agreements with these very detailed (albeit 1D and stationary) models. We also find equally good agreement with the alternative CMF line-driven wind models by \citet{krticka_2017}, which have been developed completely independently of ours and also use different line lists. Nevertheless, it will be an important task in future work to try and further improve the hybrid opacity approximation employed here.

Thirdly, in our calculations for the $\kappa^{\rm line}$
we assume the Sobolev approximation, which inevitably neglects the influence of the  LDI. While it is believed that this LDI might be strongly damped in optically thick atmospheric regions \citep{gayley_1995b}, it will likely further increase structure formation in parts of the wind that are optically thin. However, a proper inclusion of the LDI would require that we replace our formulation of $\kappa^{\rm line}$ in terms of the local velocity gradient and instead perform (very costly) non-local integrations to obtain the corresponding optical depths. Typically, all LDI simulations carried out thus far have been based on some variant of the escape-integral formalism of \citet{puls_owocki_1996}, which is different than the method presented and applied here; as such, it remains to be seen if and how our current modeling framework could be adapted to include the LDI effect. 

Fourthly, to solve the RHD equations in our simulations we use analytical closure relations between radiation energy density, pressure, and flux. 
\oc{While this analytic closure reproduces the correct optically thick and thin limits, it is again in the \q{intermediate} regime, where it may be problematic.} To evaluate the potential errors introduced by our analytical approximation, \citet{nico_2022b} calculated the radiation quantities by solving the 3D radiative transfer equation using a short characteristics (SC) method \citep{levin_2020}, which illustrated that (except in the outer winds) the analytical closure approximations overall actually yielded very similar results to such full 3D radiative transfer solutions \citep[see further discussion in][]{nico_2022b}. While this suggests that our current approximate \q{bridging laws} to close the radiation equations might not be too bad, it will nonetheless be important in future work to try and replace these (or at least further calibrate them) with full solutions to the 3D radiative transfer equation \citep[for example via a variable Eddington tensor closure, see, e.g. ][]{Jiang_2015}. \oc{How to effectively account for the accumulative effect of line-driving stemming from thousands of spectral lines within such an approach remains a challenging question for multi-dimensional simulations though.}

Finally, in this study, we have only performed 2D time-dependent simulations. Although we do not believe this affects our overall results and conclusions, some quantitative properties may change when introducing the third spatial dimension, as has been already shown in \citet{nico_2022b} for 3D WR stars. In a future study, we will extend also our O star simulations to full 3D. Similarly, in such upcoming studies, we also aim to consider the effects of rotation and magnetic fields, which are both neglected here.  

\section{Summary and outlook}
\label{summary}

In this work, we have presented the first multi-dimensional, time-dependent unified atmosphere with wind models for O stars. To this end, we use a \q{box-in-a-star} approach (including spherical correction terms), while solving the RHD equations using an analytical flux-limiting closure approximation in the radiation energy equation, and accounting properly for line-driving effects. The initial conditions for our simulations are derived using procedures very similar to those applied in the standard 1D stationary model atmosphere with wind codes; however, here these are extended to deeper layers of the stellar envelope to cover the unstable sub-surface iron-opacity peak region. Perturbations derived from linear stability analysis are then added to the 1D gas density and lateral velocity in this region to initiate structure formation. 

In our simulations, we find that structure starts appearing in the previously perturbed iron-opacity peak region. Simultaneously, around and beyond the optical photosphere, the densities are low enough so that line driving becomes efficient. This initiates a net supersonic wind outflow from the variable stellar surface, introducing more structure formation and leading to a complex interplay between the deeper atmospheric layers and the overlaying wind. As the models dynamically relax, we find a (quasi-) stable O star envelope beneath the iron opacity peak zone and a very turbulent atmosphere with wind above it. We find large turbulent velocities in the photospheric and wind layers of our simulations, in overall good qualitative agreement with results derived from spectroscopic observations of O stars.

The next part of our analysis compared the lateral and time \q{averaged} $\langle 2D \rangle$
atmospheric quantities with their 1D counterparts, focusing on the prototypical O4 model. Here, we find that envelope expansion in the 1D model is stronger than in $\langle 2D \rangle$, leading to a lower effective temperature and larger stellar radius in the former. Moreover, the density inversion present in the 1D profile gets washed away in the $\langle 2D \rangle$ profile, and the slope of the density and temperature profiles around the optical photosphere is much shallower in the $\langle 2D \rangle$ simulation (indicating a much larger characteristic photospheric scale height). As such, we introduced a simplified treatment of convection and turbulence in our 1D model, treating energy transport using standard mixing length theory, accounting for the effects of radiation pressure and cooling, and introducing a turbulent pressure term in the hydrostatic momentum equation. With this, we found good qualitative agreement in the adjusted 1D and $\langle 2D \rangle$ O4 model structures for parameters $\alpha_{\rm MLT} \sim 1.0$ and a $\varv_{\rm turb}$ that increased from zero in the lowermost parts to $\sim 90$ km/s in the photospheric layers.  The $\langle 2D \rangle$ velocity field further shows a rather complex structure in particularly the wind initiation region, going slightly negative around the photosphere before rapidly (though slightly less rapidly than in corresponding 1D stationary models) accelerating to high supersonic speeds when line-driving becomes efficient; this is due to the inverse relation between density and line-driving opacity, the outflowing wind parts also display a clear anti-correlation between density and velocity. Finally, we derived the \q{average} mass loss rates for our models and found good agreement with the detailed (albeit 1D and stationary) theoretical mass loss calculations by \citet{krticka_2017} and \citet{robin_2021}; this lends further support to our basic approach for including the effects of line-driving into our simulations.  

A key follow-up study to this work will regard the computation of synthetic spectra and spectroscopic comparison to observations. This is work underway in our group, using the 3D radiative transfer code framework by \citet{levin_2020}. Since quantities such as the mass loss rates, wind and turbulent velocities, and \q{atmospheric clumping} in our multi-dimensional simulations are emergent properties (rather than adjustable free input parameters), these studies will provide key basic tests regarding the realism of our new O star unified model atmosphere calculations. Furthermore, they will directly test whether the observed strength and broadness of photospheric and wind lines are also quantitatively matched by our models, how this may affect spectroscopic derivation of fundamental stellar parameters and, as such, provide further insights into how present-day 1D codes might have to be calibrated to account for basic multi-dimensional effects. Moreover, since the photosphere and wind are variable in our O star simulations, this will likely cause both spectral line and stochastic photometric variability (see also \citealt{schultz_2022}).    

As discussed in Sec \ref{limitations}, in another upcoming work we will also separate the flux-, Plank-, and energy-weighted mean opacities, using an approximate non-LTE method based primarily on \citet{puls_2000}. Finally, the methods developed in this paper \citep[also,][]{nico_2022b} are generally applicable also to other interesting regions of the upper Hertzsprung Russel (HR) diagram, for instance: B-supergiants and LBVs.

\begin{acknowledgements}

The computational resources used for this work were provided by Vlaams Supercomputer Centrum (VSC) funded by the Research Foundation-Flanders (FWO) and the Flemish Government. DD, CS, NM, LP, and JS acknowledge the support of the European Research Council (ERC) Horizon Europe under grant agreement number 101044048. OV, NM, LP, and JS acknowledge support from KU Leuven C1 grant MAESTRO C16/17/007, the Belgian Research Foundation Flanders (FWO) Odysseus program under grant number G0H9218N. 
FWO grant G077822N. JS would further like to extend his thanks to Stan Owocki and Jo Puls for many fruitful discussions over the years, and to the latter also for providing his easy-to-follow (albeit in hand-written German, of course) koch rezept for MLT-based convective energy transport including radiative pressure and cooling. \oc{We thank the referee for their constructive comments to the manuscript.} Finally, the authors would like to thank all members of the KUL-EQUATION group for fruitful discussion, comments, and suggestions We made significant use of the following packages to analyze our data: {\fontfamily{qcr}\selectfont NumPy} \citep{harris_2020}, {\fontfamily{qcr}\selectfont SciPy} \citep{virtanen_2020}, {\fontfamily{qcr}\selectfont matplotlib} \citep{hunter_2007}, {\fontfamily{qcr}\selectfont Python amrvac\_reader} \citep{keppens_2020}. 

\end{acknowledgements}

\bibliographystyle{aa}
\bibliography{references_ostar} 

\begin{thebibliography}{99}
\expandafter\ifx\csname natexlab\endcsname\relax\def\natexlab#1{#1}\fi

\bibitem[{{Abbott} \& {Lucy}(1985)}]{lucy_1985}
{Abbott}, D.~C. \& {Lucy}, L.~B. 1985, \apj, 288, 679

\bibitem[{{Asplund} {et~al.}(2009){Asplund}, {Grevesse}, {Sauval}, \& {Scott}}]{asplund_2009}
{Asplund}, M., {Grevesse}, N., {Sauval}, A.~J., \& {Scott}, P. 2009, \araa, 47, 481

\bibitem[{Asplund {et~al.}(2000)Asplund, Nordlund, Trampedach, Allende~Prieto, \& Stein}]{asplund_2000}
Asplund, M., Nordlund, A., Trampedach, R., Allende~Prieto, C., \& Stein, R.~F. 2000, 359, 729

\bibitem[{{Bj{\"o}rklund} {et~al.}(2021){Bj{\"o}rklund}, {Sundqvist}, {Puls}, \& {Najarro}}]{robin_2021}
{Bj{\"o}rklund}, R., {Sundqvist}, J.~O., {Puls}, J., \& {Najarro}, F. 2021, \aap, 648, A36

\bibitem[{{Blaes} \& {Socrates}(2003)}]{bs03}
{Blaes}, O. \& {Socrates}, A. 2003, \apj, 596, 509

\bibitem[{{Brands} {et~al.}(2022){Brands}, {de Koter}, {Bestenlehner}, {Crowther}, {Sundqvist}, {Puls}, {Caballero-Nieves}, {Abdul-Masih}, {Driessen}, {Garc{\'\i}a}, {Geen}, {Gr{\"a}fener}, {Hawcroft}, {Kaper}, {Keszthelyi}, {Langer}, {Sana}, {Schneider}, {Shenar}, \& {Vink}}]{brands_2022}
{Brands}, S.~A., {de Koter}, A., {Bestenlehner}, J.~M., {et~al.} 2022, \aap, 663, A36

\bibitem[{{Cantiello} {et~al.}(2009){Cantiello}, {Langer}, {Brott}, {de Koter}, {Shore}, {Vink}, {Voegler}, {Lennon}, \& {Yoon}}]{cantiello_2009}
{Cantiello}, M., {Langer}, N., {Brott}, I., {et~al.} 2009, \aap, 499, 279

\bibitem[{{Castor}(2004)}]{cak2004}
{Castor}, J.~I. 2004, {Radiation Hydrodynamics}

\bibitem[{{Castor} {et~al.}(1975){Castor}, {Abbott}, \& {Klein}}]{cak1975}
{Castor}, J.~I., {Abbott}, D.~C., \& {Klein}, R.~I. 1975, \apj, 195, 157

\bibitem[{{Conti} \& {Ebbets}(1977)}]{conti_1997}
{Conti}, P.~S. \& {Ebbets}, D. 1977, \apj, 213, 438

\bibitem[{Driessen {et~al.}(2022)Driessen, Sundqvist, \& Kee}]{flo_2022b}
Driessen, F., Sundqvist, J.~O., \& Kee, N.~D. 2022, The line-deshadowing instability and its effect on wind clumping for OB-stars

\bibitem[{{Driessen} {et~al.}(2022){Driessen}, {Sundqvist}, \& {Dagore}}]{flo_2022}
{Driessen}, F.~A., {Sundqvist}, J.~O., \& {Dagore}, A. 2022, \aap, 663, A40

\bibitem[{{Eversberg} {et~al.}(1998){Eversberg}, {L{\'e}pine}, \& {Moffat}}]{eversberg_1998}
{Eversberg}, T., {L{\'e}pine}, S., \& {Moffat}, A. F.~J. 1998, \apj, 494, 799

\bibitem[{{Freytag} {et~al.}(2012){Freytag}, {Steffen}, {Ludwig}, {Wedemeyer-B{\"o}hm}, {Schaffenberger}, \& {Steiner}}]{freytag_2012}
{Freytag}, B., {Steffen}, M., {Ludwig}, H.~G., {et~al.} 2012, Journal of Computational Physics, 231, 919

\bibitem[{{Friend} \& {Abbott}(1986)}]{friend_1986}
{Friend}, D.~B. \& {Abbott}, D.~C. 1986, \apj, 311, 701

\bibitem[{{Friend} \& {Castor}(1983)}]{castor_friend_1983}
{Friend}, D.~B. \& {Castor}, J.~I. 1983, \apj, 272, 259

\bibitem[{{Gabler} {et~al.}(1989){Gabler}, {Gabler}, {Kudritzki}, {Puls}, \& {Pauldrach}}]{gabler_1989}
{Gabler}, R., {Gabler}, A., {Kudritzki}, R.~P., {Puls}, J., \& {Pauldrach}, A. 1989, \aap, 226, 162

\bibitem[{{Gayley}(1995)}]{gayley_1995}
{Gayley}, K.~G. 1995, \apj, 454, 410

\bibitem[{{Gayley} \& {Owocki}(1995)}]{gayley_1995b}
{Gayley}, K.~G. \& {Owocki}, S.~P. 1995, \apj, 446, 801

\bibitem[{{Glatzel}(1994)}]{glatzel_1994}
{Glatzel}, W. 1994, \mnras, 271, 66

\bibitem[{Goldberg {et~al.}(2022)Goldberg, Jiang, \& Bildsten}]{goldberg_2022}
Goldberg, J.~A., Jiang, Y.-F., \& Bildsten, L. 2022, The Astrophysical Journal, 929, 156

\bibitem[{{Gr{\"a}fener} {et~al.}(2002){Gr{\"a}fener}, {Koesterke}, \& {Hamann}}]{grafener_2002}
{Gr{\"a}fener}, G., {Koesterke}, L., \& {Hamann}, W.~R. 2002, \aap, 387, 244

\bibitem[{{Gr{\"a}fener} {et~al.}(2012){Gr{\"a}fener}, {Owocki}, \& {Vink}}]{grafener_2012}
{Gr{\"a}fener}, G., {Owocki}, S.~P., \& {Vink}, J.~S. 2012, \aap, 538, A40

\bibitem[{{Grassitelli} {et~al.}(2016){Grassitelli}, {Chen{\'e}}, {Sanyal}, {Langer}, {St-Louis}, {Bestenlehner}, \& {Fossati}}]{grassitelli_2016}
{Grassitelli}, L., {Chen{\'e}}, A.~N., {Sanyal}, D., {et~al.} 2016, \aap, 590, A12

\bibitem[{{Grassitelli} {et~al.}(2018){Grassitelli}, {Langer}, {Grin}, {Mackey}, {Bestenlehner}, \& {Gr{\"a}fener}}]{grassitelli_2018}
{Grassitelli}, L., {Langer}, N., {Grin}, N.~J., {et~al.} 2018, \aap, 614, A86

\bibitem[{{Gray}(2008)}]{gray_2008}
{Gray}, D.~F. 2008, {The Observation and Analysis of Stellar Photospheres}

\bibitem[{{Grevesse} \& {Noels}(1993)}]{grevesse_1993}
{Grevesse}, N. \& {Noels}, A. 1993, in Origin and Evolution of the Elements, ed. N.~{Prantzos}, E.~{Vangioni-Flam}, \& M.~{Casse}, 15--25

\bibitem[{{Hamann} \& {Gr{\"a}fener}(2004)}]{hamann_2004}
{Hamann}, W.~R. \& {Gr{\"a}fener}, G. 2004, \aap, 427, 697

\bibitem[{{Harris} {et~al.}(2020){Harris}, {Millman}, {van der Walt}, {Gommers}, {Virtanen}, {Cournapeau}, {Wieser}, {Taylor}, {Berg}, {Smith}, {Kern}, {Picus}, {Hoyer}, {van Kerkwijk}, {Brett}, {Haldane}, {del R{\'\i}o}, {Wiebe}, {Peterson}, {G{\'e}rard-Marchant}, {Sheppard}, {Reddy}, {Weckesser}, {Abbasi}, {Gohlke}, \& {Oliphant}}]{harris_2020}
{Harris}, C.~R., {Millman}, K.~J., {van der Walt}, S.~J., {et~al.} 2020, \nat, 585, 357

\bibitem[{{Haser} {et~al.}(1995){Haser}, {Lennon}, {Kudritzki}, {Puls}, {Pauldrach}, {Bianchi}, \& {Hutchings}}]{haser_1995}
{Haser}, S.~M., {Lennon}, D.~J., {Kudritzki}, R.~P., {et~al.} 1995, \aap, 295, 136

\bibitem[{{Hawcroft} {et~al.}(2021){Hawcroft}, {Sana}, {Mahy}, {Sundqvist}, {Abdul-Masih}, {Bouret}, {Brands}, {de Koter}, {Driessen}, \& {Puls}}]{hawcroft_2021}
{Hawcroft}, C., {Sana}, H., {Mahy}, L., {et~al.} 2021, \aap, 655, A67

\bibitem[{{Hawcroft} {et~al.}(2023){Hawcroft}, {Sana}, {Mahy}, {Sundqvist}, {de Koter}, {Crowther}, {Bestenlehner}, {Brands}, {David-Uraz}, {Decin}, {Erba}, {Garcia}, {Hamann}, {Herrero}, {Ignace}, {Kee}, {Kub{\'a}tov{\'a}}, {Lefever}, {Moffat}, {Najarro}, {Oskinova}, {Pauli}, {Prinja}, {Puls}, {Sander}, {Shenar}, {St-Louis}, {ud-Doula}, \& {Vink}}]{hawcroft_2023}
{Hawcroft}, C., {Sana}, H., {Mahy}, L., {et~al.} 2023, arXiv e-prints, arXiv:2303.12165

\bibitem[{{Hearn}(1972)}]{hearn_1972}
{Hearn}, A.~G. 1972, \aap, 19, 417

\bibitem[{{Hennicker} {et~al.}(2020){Hennicker}, {Puls}, {Kee}, \& {Sundqvist}}]{levin_2020}
{Hennicker}, L., {Puls}, J., {Kee}, N.~D., \& {Sundqvist}, J.~O. 2020, \aap, 633, A16

\bibitem[{{Herrero} {et~al.}(1992){Herrero}, {Kudritzki}, {Vilchez}, {Kunze}, {Butler}, \& {Haser}}]{herrero_1992}
{Herrero}, A., {Kudritzki}, R.~P., {Vilchez}, J.~M., {et~al.} 1992, \aap, 261, 209

\bibitem[{{Hillier}(1991)}]{hillier_1991}
{Hillier}, D.~J. 1991, \aap, 247, 455

\bibitem[{{Hillier} \& {Lanz}(2001)}]{hillier_2001}
{Hillier}, D.~J. \& {Lanz}, T. 2001, in Astronomical Society of the Pacific Conference Series, Vol. 247, Spectroscopic Challenges of Photoionized Plasmas, ed. G.~{Ferland} \& D.~W. {Savin}, 343

\bibitem[{{Hillier} \& {Miller}(1998)}]{hillier_1998}
{Hillier}, D.~J. \& {Miller}, D.~L. 1998, \apj, 496, 407

\bibitem[{Howarth {et~al.}(1997)Howarth, Siebert, Hussain, \& Prinja}]{howarth_1997}
Howarth, I.~D., Siebert, K.~W., Hussain, G. A.~J., \& Prinja, R.~K. 1997, Monthly Notices of the Royal Astronomical Society, 284, 265

\bibitem[{{Hunter}(2007)}]{hunter_2007}
{Hunter}, J.~D. 2007, Computing in Science and Engineering, 9, 90

\bibitem[{{Iglesias} \& {Rogers}(1996)}]{opal}
{Iglesias}, C.~A. \& {Rogers}, F.~J. 1996, \apj, 464, 943

\bibitem[{{Jiang}(2023)}]{jiang_2023}
{Jiang}, Y.-F. 2023, Galaxies, 11, 105

\bibitem[{{Jiang} {et~al.}(2015){Jiang}, {Cantiello}, {Bildsten}, {Quataert}, \& {Blaes}}]{Jiang_2015}
{Jiang}, Y.-F., {Cantiello}, M., {Bildsten}, L., {Quataert}, E., \& {Blaes}, O. 2015, \apj, 813, 74

\bibitem[{{Jiang} {et~al.}(2018){Jiang}, {Cantiello}, {Bildsten}, {Quataert}, {Blaes}, \& {Stone}}]{jiang_2018}
{Jiang}, Y.-F., {Cantiello}, M., {Bildsten}, L., {et~al.} 2018, \nat, 561, 498

\bibitem[{{Karp} {et~al.}(1977){Karp}, {Lasher}, {Chan}, \& {Salpeter}}]{karp_lasher}
{Karp}, A.~H., {Lasher}, G., {Chan}, K.~L., \& {Salpeter}, E.~E. 1977, \apj, 214, 161

\bibitem[{{Kee} {et~al.}(2016){Kee}, {Owocki}, \& {Sundqvist}}]{kee_2016}
{Kee}, N.~D., {Owocki}, S., \& {Sundqvist}, J.~O. 2016, \mnras, 458, 2323

\bibitem[{{Keppens} {et~al.}(2023){Keppens}, {Popescu Braileanu}, {Zhou}, {Ruan}, {Xia}, {Guo}, {Claes}, \& {Bacchini}}]{rony_2023}
{Keppens}, R., {Popescu Braileanu}, B., {Zhou}, Y., {et~al.} 2023, \aap, 673, A66

\bibitem[{{Keppens} {et~al.}(2020){Keppens}, {Teunissen}, {Xia}, \& {Porth}}]{keppens_2020}
{Keppens}, R., {Teunissen}, J., {Xia}, C., \& {Porth}, O. 2020, arXiv e-prints, arXiv:2004.03275

\bibitem[{{Kippenhahn} {et~al.}(2013){Kippenhahn}, {Weigert}, \& {Weiss}}]{kip_2013}
{Kippenhahn}, R., {Weigert}, A., \& {Weiss}, A. 2013, {Stellar Structure and Evolution}

\bibitem[{{K{\"o}hler} {et~al.}(2015){K{\"o}hler}, {Langer}, {de Koter}, {de Mink}, {Crowther}, {Evans}, {Gr{\"a}fener}, {Sana}, {Sanyal}, {Schneider}, \& {Vink}}]{kohler_2015}
{K{\"o}hler}, K., {Langer}, N., {de Koter}, A., {et~al.} 2015, \aap, 573, A71

\bibitem[{{Krti{\v{c}}ka} \& {Kub{\'a}t}(2017)}]{krticka_2017}
{Krti{\v{c}}ka}, J. \& {Kub{\'a}t}, J. 2017, \aap, 606, A31

\bibitem[{{Lucy}(1971)}]{lucy_1971}
{Lucy}, L.~B. 1971, \apj, 163, 95

\bibitem[{{Lucy} \& {Abbott}(1993)}]{lucy_1993}
{Lucy}, L.~B. \& {Abbott}, D.~C. 1993, \apj, 405, 738

\bibitem[{{Mihalas} \& {Mihalas}(1984)}]{mihalas_1984}
{Mihalas}, D. \& {Mihalas}, B.~W. 1984, {Foundations of radiation hydrodynamics}

\bibitem[{{Moens} {et~al.}(2022{\natexlab{a}}){Moens}, {Poniatowski}, {Hennicker}, {Sundqvist}, {El Mellah}, \& {Kee}}]{nico_2022b}
{Moens}, N., {Poniatowski}, L.~G., {Hennicker}, L., {et~al.} 2022{\natexlab{a}}, \aap, 665, A42

\bibitem[{{Moens} {et~al.}(2022{\natexlab{b}}){Moens}, {Sundqvist}, {El Mellah}, {Poniatowski}, {Teunissen}, \& {Keppens}}]{nico_2022a}
{Moens}, N., {Sundqvist}, J.~O., {El Mellah}, I., {et~al.} 2022{\natexlab{b}}, \aap, 657, A81

\bibitem[{{Oskinova} {et~al.}(2007){Oskinova}, {Hamann}, \& {Feldmeier}}]{oskinova_2007}
{Oskinova}, L.~M., {Hamann}, W.~R., \& {Feldmeier}, A. 2007, \aap, 476, 1331

\bibitem[{{Owocki}(2015)}]{owocki_2015}
{Owocki}, S.~P. 2015, in Astrophysics and Space Science Library, Vol. 412, Very Massive Stars in the Local Universe, ed. J.~S. {Vink}, 113

\bibitem[{{Owocki} {et~al.}(1988){Owocki}, {Castor}, \& {Rybicki}}]{owocki_1998}
{Owocki}, S.~P., {Castor}, J.~I., \& {Rybicki}, G.~B. 1988, \apj, 335, 914

\bibitem[{{Owocki} {et~al.}(2004){Owocki}, {Gayley}, \& {Shaviv}}]{owocki_2004}
{Owocki}, S.~P., {Gayley}, K.~G., \& {Shaviv}, N.~J. 2004, \apj, 616, 525

\bibitem[{{Owocki} \& {Puls}(1996)}]{puls_owocki_1996}
{Owocki}, S.~P. \& {Puls}, J. 1996, \apj, 462, 894

\bibitem[{{Owocki} \& {Rybicki}(1984)}]{owocki_1984}
{Owocki}, S.~P. \& {Rybicki}, G.~B. 1984, \apj, 284, 337

\bibitem[{{Owocki} \& {Sundqvist}(2018)}]{owocki_sund_2018}
{Owocki}, S.~P. \& {Sundqvist}, J.~O. 2018, \mnras, 475, 814

\bibitem[{{Owocki} {et~al.}(2017){Owocki}, {Townsend}, \& {Quataert}}]{owocki_2017}
{Owocki}, S.~P., {Townsend}, R. H.~D., \& {Quataert}, E. 2017, \mnras, 472, 3749

\bibitem[{{Pauldrach} {et~al.}(1986){Pauldrach}, {Puls}, \& {Kudritzki}}]{pauldrach_1986}
{Pauldrach}, A., {Puls}, J., \& {Kudritzki}, R.~P. 1986, \aap, 164, 86

\bibitem[{{Pauldrach} {et~al.}(2001){Pauldrach}, {Hoffmann}, \& {Lennon}}]{pauldrach_2001}
{Pauldrach}, A.~W.~A., {Hoffmann}, T.~L., \& {Lennon}, M. 2001, \aap, 375, 161

\bibitem[{{Pauldrach} {et~al.}(1998){Pauldrach}, {Lennon}, {Hoffmann}, {Sellmaier}, {Kudritzki}, \& {Puls}}]{pauldrach_1998}
{Pauldrach}, A.~W.~A., {Lennon}, M., {Hoffmann}, T.~L., {et~al.} 1998, in Astronomical Society of the Pacific Conference Series, Vol. 131, Properties of Hot Luminous Stars, ed. I.~{Howarth}, 258

\bibitem[{{Petrenz} \& {Puls}(2000)}]{petrenz_puls_2000}
{Petrenz}, P. \& {Puls}, J. 2000, in Astronomical Society of the Pacific Conference Series, Vol. 214, IAU Colloq. 175: The Be Phenomenon in Early-Type Stars, ed. M.~A. {Smith}, H.~F. {Henrichs}, \& J.~{Fabregat}, 626

\bibitem[{{Petrovic} {et~al.}(2006){Petrovic}, {Pols}, \& {Langer}}]{petrovic_2006}
{Petrovic}, J., {Pols}, O., \& {Langer}, N. 2006, \aap, 450, 219

\bibitem[{{Poniatowski} {et~al.}(2022){Poniatowski}, {Kee}, {Sundqvist}, {Driessen}, {Moens}, {Owocki}, {Gayley}, {Decin}, {de Koter}, \& {Sana}}]{luka_2022}
{Poniatowski}, L.~G., {Kee}, N.~D., {Sundqvist}, J.~O., {et~al.} 2022, \aap, 667, A113

\bibitem[{{Poniatowski} {et~al.}(2021){Poniatowski}, {Sundqvist}, {Kee}, {Owocki}, {Marchant}, {Decin}, {de Koter}, {Mahy}, \& {Sana}}]{luka_2021}
{Poniatowski}, L.~G., {Sundqvist}, J.~O., {Kee}, N.~D., {et~al.} 2021, \aap, 647, A151

\bibitem[{Puls {et~al.}(2006)Puls, Markova, Scuderi, Stanghellini, Taranova, Burnley, \& Howarth}]{puls_2006}
Puls, J., Markova, N., Scuderi, S., {et~al.} 2006, Astronomy \& Astrophysics, 454, 625

\bibitem[{Puls {et~al.}(2020)Puls, Najarro, Sundqvist, \& Sen}]{puls_2020}
Puls, J., Najarro, F., Sundqvist, J.~O., \& Sen, K. 2020, Astronomy \& Astrophysics, 642, A172

\bibitem[{{Puls} {et~al.}(2000){Puls}, {Springmann}, \& {Lennon}}]{puls_2000}
{Puls}, J., {Springmann}, U., \& {Lennon}, M. 2000, \aaps, 141, 23

\bibitem[{{Puls} {et~al.}(2005){Puls}, {Urbaneja}, {Venero}, {Repolust}, {Springmann}, {Jokuthy}, \& {Mokiem}}]{puls_2005}
{Puls}, J., {Urbaneja}, M.~A., {Venero}, R., {et~al.} 2005, \aap, 435, 669

\bibitem[{{Rogers} \& {Iglesias}(1992)}]{iglesias_1992}
{Rogers}, F.~J. \& {Iglesias}, C.~A. 1992, \apjs, 79, 507

\bibitem[{{Rybicki} {et~al.}(1990){Rybicki}, {Owocki}, \& {Castor}}]{rybicki_1990}
{Rybicki}, G.~B., {Owocki}, S.~P., \& {Castor}, J.~I. 1990, \apj, 349, 274

\bibitem[{Sander {et~al.}(2012)Sander, Hamann, \& Todt}]{sander_2012}
Sander, A., Hamann, W.-R., \& Todt, H. 2012, Astronomy \& Astrophysics, 540, A144

\bibitem[{{Sander} {et~al.}(2017){Sander}, {Hamann}, {Todt}, {Hainich}, \& {Shenar}}]{sander_2017}
{Sander}, A.~A.~C., {Hamann}, W.~R., {Todt}, H., {Hainich}, R., \& {Shenar}, T. 2017, \aap, 603, A86

\bibitem[{{Santolaya-Rey} {et~al.}(1997){Santolaya-Rey}, {Puls}, \& {Herrero}}]{santolaya}
{Santolaya-Rey}, A.~E., {Puls}, J., \& {Herrero}, A. 1997, \aap, 323, 488

\bibitem[{{Sanyal} {et~al.}(2015){Sanyal}, {Grassitelli}, {Langer}, \& {Bestenlehner}}]{sanyal_2015}
{Sanyal}, D., {Grassitelli}, L., {Langer}, N., \& {Bestenlehner}, J.~M. 2015, \aap, 580, A20

\bibitem[{{Schultz} {et~al.}(2020){Schultz}, {Bildsten}, \& {Jiang}}]{schultz_2020}
{Schultz}, W.~C., {Bildsten}, L., \& {Jiang}, Y.-F. 2020, \apj, 902, 67

\bibitem[{Schultz {et~al.}(2022)Schultz, Bildsten, \& Jiang}]{schultz_2022}
Schultz, W.~C., Bildsten, L., \& Jiang, Y.-F. 2022, The Astrophysical Journal Letters, 924, L11

\bibitem[{{Schwarzschild}(1958)}]{schwarzschild_1958}
{Schwarzschild}, M. 1958, {Structure and evolution of the stars.}

\bibitem[{{Shaviv}(1998)}]{shaviv_1998}
{Shaviv}, N.~J. 1998, \apjl, 494, L193

\bibitem[{{Sim{\'o}n-D{\'\i}az} {et~al.}(2017){Sim{\'o}n-D{\'\i}az}, {Godart}, {Castro}, {Herrero}, {Aerts}, {Puls}, {Telting}, \& {Grassitelli}}]{simon_2017}
{Sim{\'o}n-D{\'\i}az}, S., {Godart}, M., {Castro}, N., {et~al.} 2017, \aap, 597, A22

\bibitem[{{Sobolev}(1960)}]{sobolev_1960}
{Sobolev}, V.~V. 1960, {Moving Envelopes of Stars}

\bibitem[{{Stothers} \& {Chin}(1993)}]{slothers_1993}
{Stothers}, R.~B. \& {Chin}, C.-W. 1993, \apj, 412, 294

\bibitem[{{Sundqvist} {et~al.}(2019){Sundqvist}, {Bj{\"o}rklund}, {Puls}, \& {Najarro}}]{sundqvist_bjorklund_2019}
{Sundqvist}, J.~O., {Bj{\"o}rklund}, R., {Puls}, J., \& {Najarro}, F. 2019, \aap, 632, A126

\bibitem[{{Sundqvist} {et~al.}(2018){Sundqvist}, {Owocki}, \& {Puls}}]{sundqvist_2018}
{Sundqvist}, J.~O., {Owocki}, S.~P., \& {Puls}, J. 2018, \aap, 611, A17

\bibitem[{{Sundqvist} \& {Puls}(2018)}]{sundqvist_2018b}
{Sundqvist}, J.~O. \& {Puls}, J. 2018, \aap, 619, A59

\bibitem[{{Sundqvist} {et~al.}(2010){Sundqvist}, {Puls}, \& {Feldmeier}}]{sundqvist_2010}
{Sundqvist}, J.~O., {Puls}, J., \& {Feldmeier}, A. 2010, \aap, 510, A11

\bibitem[{{Sundqvist} {et~al.}(2013){Sundqvist}, {Sim{\'o}n-D{\'\i}az}, {Puls}, \& {Markova}}]{sundqvist_2013}
{Sundqvist}, J.~O., {Sim{\'o}n-D{\'\i}az}, S., {Puls}, J., \& {Markova}, N. 2013, \aap, 559, L10

\bibitem[{{Teunissen} \& {Keppens}(2019)}]{janis_rony_19}
{Teunissen}, J. \& {Keppens}, R. 2019, Computer Physics Communications, 245, 106866

\bibitem[{Turner \& Stone(2001)}]{turner_2001}
Turner, N.~J. \& Stone, J.~M. 2001, The Astrophysical Journal Supplement Series, 135, 95

\bibitem[{ud~Doula \& Owocki(2002)}]{ud_doula_2002}
ud~Doula, A. \& Owocki, S.~P. 2002, The Astrophysical Journal, 576, 413

\bibitem[{{Virtanen} {et~al.}(2020){Virtanen}, {Gommers}, {Oliphant}, {Haberland}, {Reddy}, {Cournapeau}, {Burovski}, {Peterson}, {Weckesser}, {Bright}, {van der Walt}, {Brett}, {Wilson}, {Millman}, {Mayorov}, {Nelson}, {Jones}, {Kern}, {Larson}, {Carey}, {Polat}, {Feng}, {Moore}, {VanderPlas}, {Laxalde}, {Perktold}, {Cimrman}, {Henriksen}, {Quintero}, {Harris}, {Archibald}, {Ribeiro}, {Pedregosa}, {van Mulbregt}, \& {SciPy 1. 0 Contributors}}]{virtanen_2020}
{Virtanen}, P., {Gommers}, R., {Oliphant}, T.~E., {et~al.} 2020, Nature Methods, 17, 261

\bibitem[{{{\v{S}}urlan} {et~al.}(2012){{\v{S}}urlan}, {Hamann}, {Kub{\'a}t}, {Oskinova}, \& {Feldmeier}}]{surlan_2012}
{{\v{S}}urlan}, B., {Hamann}, W.~R., {Kub{\'a}t}, J., {Oskinova}, L.~M., \& {Feldmeier}, A. 2012, \aap, 541, A37

\bibitem[{Xia {et~al.}(2018)Xia, Teunissen, Mellah, Chané, \& Keppens}]{xia_2018}
Xia, C., Teunissen, J., Mellah, I.~E., Chané, E., \& Keppens, R. 2018, The Astrophysical Journal Supplement Series, 234, 30

\end{thebibliography}

\begin{appendix}

\section{Castor's expansion opacity}
\label{appendix_a}

The expansion opacity model uses the escape probability concept from \citet{sobolev_1960} in a situation when a spectrum is populated with multiple lines and the medium has a significant velocity gradient (\citealt{karp_lasher,cak2004}). Under these conditions, for lines with a frequency separation corresponding to a velocity interval $\Delta \varv < \varv_\infty$, the photon mean-free-path between interactions with different lines is $\ell_{\rm mfp} = 1/(\kappa_{\rm eff} \rho)$, where $\kappa_{\rm eff}$ is the effective opacity. 
Here $l_{\rm mfp} = \Delta l/P$ for $\Delta l$ the physical mean distance between two lines and $P = 1 - e^{- \tau_{s}}$ the line transition probability. By means of the Doppler shift and Sobolev line optical depth, we obtain:   
\begin{align}
    & \tau_{s} = \frac{\alpha_{l} c }{\nu_0} \left|{\frac{d\varv_{l}}{dl}}\right|^{-1},
    \label{sob_app}
\end{align}
and the effective line opacity is thus:   
\begin{align}
    & \kappa_{\rm eff} = \frac{1}{\rho} \frac{\alpha_{l}}{\Delta \nu} \frac{1 - e^{-\tau_{s}}}{\tau_{s}}, 
\end{align}
where $\alpha_l$ is the frequency-integrated line extinction coefficient. Considering a frequency interval $\Delta \nu$ within which there is \textit{i} number of lines (in a line list), and adding a continuum opacity source given by Thompson scattering $\kappa_{\rm TH}$, the total effective opacity is 
\begin{align}
    & \kappa_{\rm eff} = \frac{1}{\rho} \sum_{i}^{\rm lines \ in  \ \Delta \nu} \frac{\alpha_{l}^i}{\Delta \nu} \left(\frac{1 - e^{-\tau_s^i}}{\tau_s^i}\right) + \kappa_{\rm TH} \\
    & = \frac{d \varv_{l}}{dl} \frac{1}{c \rho} \sum_{i}^{\rm lines \ in  \ \Delta \nu} \frac{\nu^i}{\Delta \nu} \left(1 - e^{-\tau_s^i}\right) + \kappa_{\rm TH}.
    \label{kappa_eff}
\end{align}
The latter expression is equivalent to \citet{cak2004}'s Eq. 6.130 (also Eq. 9 in \citealt{castor_friend_1983}) for the effective opacity. 

Furthermore, \citet{cak2004} performs explicit opacity calculations using a Fe III line list in LTE (his Sect. 6.9.4). He finds that for large values of $\tau = \tau_S / q_i$ ($\tau = s$ in his notation), the above formula considerably underestimates the effective opacity since the Sobolev line opacity tends to zero when the velocity gradient becomes very small, due to the neglect of the intrinsic line width. To make the expansion opacity formula agree better with explicit calculations, thus, the author suggests that the Thomson continuum be replaced with the actual harmonic Rosseland mean opacity calculated without the velocity gradient (but including the intrinsic line widths); namely,    
\begin{align}
    & \kappa_{\rm eff} = \frac{1}{\rho} \sum_{i}^{\rm lines \ in  \ \Delta \nu} \frac{\alpha_{l}^i}{\Delta \nu} \left(\frac{1 - e^{-\tau_s^i}}{\tau_s^i}\right) + \kappa^{\rm Ross}
    \label{kappa_eff_ross}
\end{align}
\paragraph{Connection to hybrid opacity model}
The flux-weighted opacity of the above expansion opacity is:  
\begin{align}
    \frac{\kappa_{\rm eff} F_\nu \Delta\nu}{F}
\end{align}
when taken over the complete frequency range, this becomes: 
\begin{align}
  & \kappa  = \kappa_0 \sum_{i}^{\rm all \ lines} q_i w_i \left(\frac{1 - e^{-\tau_s^i}}{\tau_s^i}\right) + \kappa^{\rm Ross}   
\end{align}
which we have translated to notation used in the present paper; this illustrates how the hybrid opacity model suggested by \citet{luka_2022} in effect is equivalent to the suggestion in \citet{cak2004} for modification of the Sobolev expansion opacity to account for the intrinsic widths of the lines in the limit of large values for $\tau \, (\propto \rho/d \varv_l /dl)$.  

\end{appendix}

\end{document}